\documentclass[journal,10pt]{IEEEtran}

\usepackage{times,epsfig,latexsym,multirow,array,amssymb,graphicx,multicol,stfloats,textcomp,psfrag,bm}
\usepackage[cmex10]{amsmath}
\usepackage{eqparbox}
\usepackage{xcolor}
\usepackage{cite}
\usepackage{algorithm}
\usepackage{algpseudocode}

\begin{document}
\title{Integrating PHY Security Into NDN-IoT Networks By Exploiting MEC: Authentication Efficiency, Robustness, and Accuracy Enhancement}

\author{
Peng~Hao,
       and~Xianbin~Wang,~\IEEEmembership{Fellow,~IEEE}
\thanks{

The authors are with the Department of Electrical and Computer Engineering,
Western University, London, ON, N6A 5B9, Canada (e-mail: phao5@uwo.ca; xianbin.wang@uwo.ca)}
}

\maketitle

\IEEEpeerreviewmaketitle
\begin{abstract}

Recent literature has demonstrated the improved data discovery and delivery efficiency gained through applying named data networking (NDN) to a variety of information-centric Internet of things (IoT) applications.
However, from a data security perspective, the development of NDN-IoT raises several new authentication challenges. In particular, NDN-IoT authentication may require per-packet-level signatures, thus imposing intolerably high computational and time costs on the resource-poor IoT end devices.
This paper proposes an effective solution by seamlessly integrating the lightweight and unforgeable physical-layer identity (PHY-ID) into the existing NDN signature scheme for the mobile edge computing (MEC)-enabled NDN-IoT networks. 
The PHY-ID generation exploits the inherent signal-level device-specific radio-frequency imperfections of IoT devices, including the in-phase/quadrature-phase imbalance, and thereby avoids adding any implementation complexity to the constrained IoT devices. We derive the offline maximum entropy-based quantization rule and propose an online two-step authentication scheme to improve the accuracy of the authentication decision making. Consequently, a cooperative MEC device can securely execute the costly signing task on behalf of the authenticated IoT device in an optimal manner. The evaluation results demonstrate 1) elevated authentication time efficiency, 2) robustness to several impersonation attacks including the replay attack and the computation-based spoofing attack, and 3) increased differentiation rate and correct authentication probability by applying our integration design in MEC-enabled NDN-IoT networks.

\end{abstract}

\section{Introduction}

\IEEEPARstart{T}{o} build secure, scalable smart city networks, an enormous variety of new Internet of things (IoT) devices are deployed ubiquitously every day to sense every concern of the city with minimal human intervention and to provide the sensed data to various information-centric applications via the Internet. As part of efforts to supportively address the explosive increase in IoT data traffic, investigations of mobile edge computing (MEC) architecture and named data networking (NDN) have aroused extensive attention in recent years. The MEC-enabled IoT can supplement the cloud computing-based IoT through harvesting the cache/storage ability, the idle computational capacity, and the context-aware resources located at the network edge \cite{7727082}. NDN with the abilities of data content naming\footnote{In NDN-IoT, the information produced by the IoT end devices at the network edge can be regarded as content \cite{NDN_IoT}.} and in-network caching can be beneficial to the IoT in terms of supporting device mobility \cite{contentRetrievalMobile} and raising the efficiency of data discovery and delivery \cite{NDN_IoT} for the information-centric applications, such as intelligent transportation systems \cite{NDNsmartcity,ITSNDN} and building automation and management systems \cite{6843232}. These IoT applications need to monitor and process big volumes of sensitive information collected from a large number of simple sensors. Given the open nature of wireless propagation and the constrained computational power of wireless IoT end devices (EDs), EDs are especially vulnerable to the impersonation-based DDoS attacks, e.g., content poisoning \cite{6614127}. From the data security perspective, authentication is of significant importance either before granting any connectivity of IoT devices to the NDN or accepting any contents from the NDN data producers.

\subsection{Research Motivations}

In most host-centric IoT applications, data provenance is guaranteed by authenticating an instantaneous end-to-end  (E2E) communication channel/session. While the information-centric NDN-IoT applications need to validate the data packets that are fetched from the intermediate caches \cite{NDN_IoT,ITSNDN}. In this regard, the authentication task of NDN has to tie to the content of every data packet directly rather than securing the E2E connection. Currently, NDN relies on the digital signature to protect the data provenance at the network layer, which, most of the time, requires the data producer (i.e., IoT end devices) to execute asymmetric cryptography at a packet granularity \cite{NDN,NDNsecurity2,NDNsmartcity,ICNSurvey}. The workload of the signing task can be intolerably high, especially for low-end sensor-based IoT devices \cite{6882665,SurveyMEC}. Therefore, offloading such heavy cryptographic workloads from constrained IoT devices is promising for implementing efficient and secure NDN-IoT networks, motivating the exploitation of the two following helpful techniques. 

\textit{Per-signal-level PHY RF fingerprinting}: 
At a wireless IoT transmitter, every encapsulated NDN packet will be sent to the physical layer and radiated through the RF front-end and is thus inevitably affected by the RF imperfection parameters. At the receiver, these device-specific parameters can be extracted from any NDN packet and can represent the unique fingerprint/identity of the ED. Moreover, the estimation of RF parameters is usually a mandatory function of a receiver for signal compensation purposes. These readily available estimates can be cost-effectively used to produce an unforgeable PHY identity (PHY-ID) for transmitter authentication \cite{reportIQI5,noncrypto,IoTauthen1,IQIAFproof,imperfection1}, which has the potential to help relieve the excessive computational burden of the signature encryption.

\textit{Exploitation of MEC}: 
MEC devices (MECDs) located at the network edge can quickly observe/collect the PHY information of the served IoT devices, including the RF fingerprints, and usually possess more powerful CPUs than the EDs for performing encryption. The NDN naming and in-network caching schemes with built-in interest/data primitives are well-suited for MEC-enabled networks \cite{NDNforMEC,NDNauthen1}.

Consequently, it is practically feasible to replace the NDN packet signing at the resource-constrained ED with two steps: 1) MECD authenticates ED using PHY-ID, and 2) MECD works with the authenticated EDs in a coordinated manner to offload the signing task from EDs \cite{offload1,offload2,offload3}.

\subsection{Related Work and Technical Problems}

In the field of NDN signature schemes, the US-founded NDN project\cite{NDN} defined the architecture and functions of packet signature-based authentication. The detailed,  usable NDN signature scheme was introduced in \cite{thesisNDN}. The work in \cite{NDNsecurity2} and \cite{6882665} reported the high-cost problem caused by executing signature-based authentication at simple IoT sensors without providing detailed solutions. An encryption authentication protocol in which the data provider and consumer should directly find each other with agreed services at the mobile edge was designed in \cite{NDNauthen1}. The studies of \cite{NDNauthen2} and \cite{NDNauthen3} integrated identity-based cryptography (IBC) and hierarchical IBC into NDN for authentication purposes. In a recent study \cite{NDNsmartcity}, a symmetric key-based IoT device (standard node) authentication along with NDN routing was proposed for the smart city scenario. However, applying these methods in NDN-IoT could require additional encryptions to be carried out by the constrained IoT devices.  Most importantly, the inherent problem of the above-mentioned cryptography-based authentication methods is that the impersonation attackers with rapidly growing processing power are capable of compromising or even reconstructing the encryption credentials (e.g., key and certificate) of a resource-limited IoT device in a much shorter time than before \cite{IQIAFproof}, e.g., via brute-force computation. The encryption credentials are essentially virtual numbers that are not directly associated with the hardware of IoT devices. In this case, an attacker possessing the correct credentials could be authenticated and can thus inject poisoned contents into NDN without triggering any alarm. However, due to the limited memory, computing, and energy capacities of IoT data providers, it is impractical to prevent the computation-based malicious cryptanalysis by straightforwardly increasing the computational intractability of the NDN-IoT system, e.g., using a longer key.

In this case, \cite{NDN} highlighted the prospects of investigating a secure collaboration scheme that allows trusted proxies to sign on behalf of the constrained devices. The authors of \cite{6843232} considered using the gateway to sign packets on behalf of the ED in building management systems. In \cite{offload1,offload2,offload3}, several collaborative task offloading schemes that exploit the CPU resources between MECDs and end users were investigated. Given that the signature in NDN packet is used for data provenance verification, the authenticity of ``signing task provider'' should be verified before the MECD accepts the unsigned packets, but with two requirements: 1) the authentication cannot impose much additional computational burden on resource-constrained EDs, and 2) the ED authentication can be completed even at the packet granularity.

Signal-level PHY RF fingerprinting can meet these two requirements and can be used in the MEC-enabled NDN-IoT networks as mentioned earlier. In \cite{devicefingerprint}, RF-based device fingerprinting was surveyed, and the implementation of modulation domain in-phase/quadrature-phase imbalance (IQI)-based fingerprinting was reported to be less complicated than that of the waveform domain methods. In \cite{reportIQI5,reportIQI6,reportIQI1,reportIQI4}, unclonable IQI fingerprints were used to authenticate wireless transmitters in different device-to-device scenarios. The studies of \cite{IQIAFproof,AccessPeng} proved that the stable RF-based PHY-ID is suitable for integration into cryptography primitives for the dynamically fluid IoT environment. The well-established RF fingerprinting technology is robust in nature due to the immense difficulty in arbitrarily changing hardware-level RF features within the short duration of an authentication session. 
Additionally, given the low distinguishability of range-limited RF parameters, the authors of \cite{AccessPeng} used the entropy to measure the randomness of PHY-ID and proposed the multiple-attributes multiple-observations (MAMO) technique to improve the authentication accuracy through increasing the entropy. However, the optimal entropy-based method was not derived, and thus its performance under a high density of IoT devices is obviously deteriorated. To the best of the authors' knowledge, the practical exploitation of PHY fingerprinting and MEC architecture for NDN-IoT authentication still faces two technical challenges.

\textbf{Challenge 1}: A design is needed to integrate PHY fingerprinting into NDN-IoT signature scheme seamlessly in order to accomplish secure signing task offloading.

\textbf{Challenge 2}: Achieving high authentication accuracy and strong robustness in a high-density IoT environment.

\subsection{Our Contributions}
This paper investigates the integration of PHY fingerprinting into the NDN-IoT signature scheme to achieve the signing task offloading and address the above challenges. We consider an MEC-enabled NDN-IoT network in which groups of 
IoT EDs can serve as the data providers, an MEC device can be a small-scale data center co-located with the gateway that connects IoT devices and the NDN, and NDN is capable of preliminarily processing the received NDN interest/data packets. In an offline phase, the MECD can prepare the maximum entropy-based (MEB) quantization rules and register the legitimate PHY-IDs, which are associated with the RF IQI of the EDs. By integrating this PHY-ID into the NDN-IoT signature scheme, the MECD can examine packets provenance using our two-step authentication, securely execute the costly signing task on behalf of the authenticated IoT device in an optimal manner, and finally publish the contents under its application-defined namespace.
The technical contributions of this paper are summarized as follows.

\begin{itemize} 

\item We complete the integration design with limited computational complexity imposed on the constrained IoT device. As a result, the packet signing task can be securely offloaded from the constrained IoT devices, which improves the time efficiency of the NDN-IoT network.

\item Given that diverse estimation methods can produce different IQI parameters, we derive the MEB quantization rule to generally address various IQI parameters rather than merely a specific one. Using our offline MEB quantizer and online two-step authentication, the authentication accuracy is improved in terms of the increased differentiation rate and correct authentication probability.

\item Compared to traditional encryption-based security, our PHY-aided authentication exhibits stronger resistance to an impersonation attacker with the compromised signing key and thus can effectively prevent content poisoning DDoS attacks in NDN-IoT networks.
\end{itemize}

The remainder of this paper is organized as follows. In Section \uppercase\expandafter{\romannumeral2}, the system framework and the proposed method are described. The proposed offline MEB quantization rule, integration design, online two-step authentication, and signing task offloading are described in Section \ref{MEBsection}, Section \ref{integrationsection}, Section \ref{twostep}, and Section \ref{SectionAlgorithm}, respectively. Section \ref{evaluationsection} evaluates the performance of the proposed method. The conclusions are presented in Section \ref{conclusionsection}.

\emph{Notations}: $(\cdot)^*$, $\lfloor\cdot\rfloor$, $\lceil \cdot \rceil$, and $(\cdot)^{T}$ denote conjugate, floor function, ceiling function, and transpose operations, respectively. Bold lowercase and uppercase letters represent vectors and matrices, respectively. $a_{(i)}$ denotes the $i$th element of vector $\mathbf a$. $\det(\mathbf A)$ is the determinant of matrix $\mathbf A$. $\Re(x)$ and $\Im(x)$ denote the real part and imaginary part of $x$, respectively. $\max(\cdot)$ and $\min(\cdot)$ return the maximum value and minimum value.

\section{Framework of The System And Proposed Method}\label{systemsection}

\subsection{Overview of the MEC-enabled NDN-IoT Network and RF Signal Model}
\begin{figure}[!t]
\centering
\includegraphics[width=78mm]{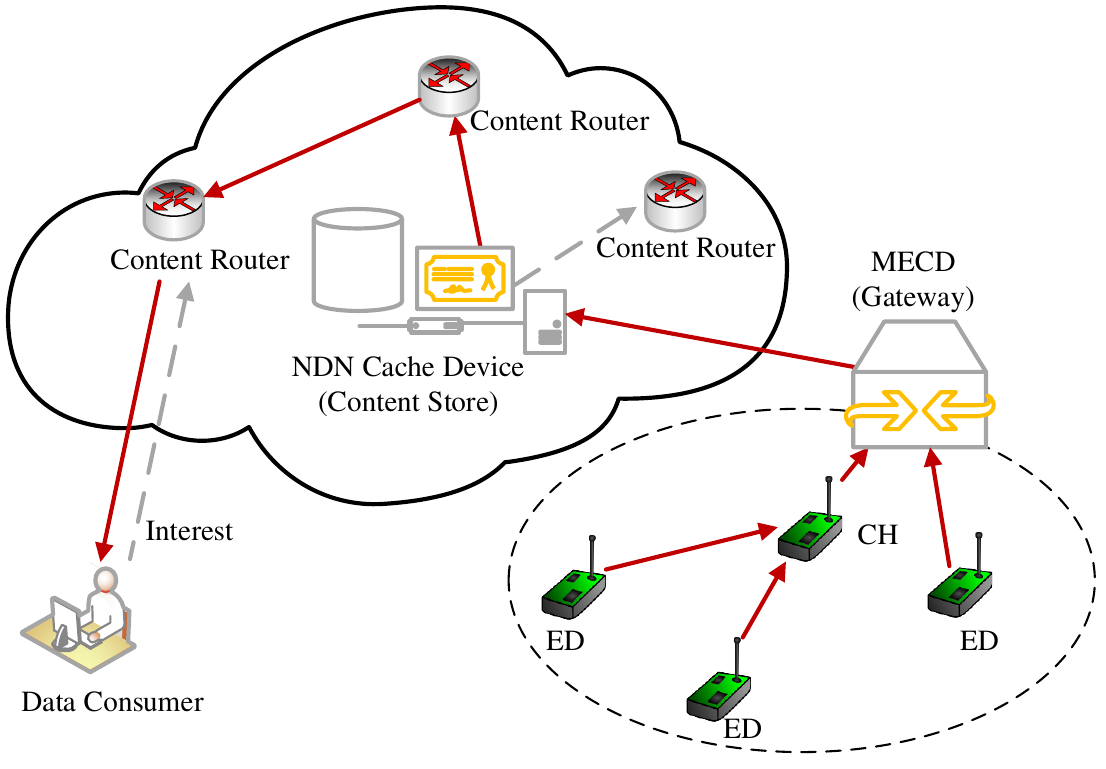}
\caption{Network structure of the MEC-enabled NDN-IoT system.}
\label{system_model}
\end{figure}

Fig.~\ref{system_model} illustrates the considered MEC-enabled NDN-IoT network. A cluster of IoT EDs can serve as the data provider to send data to consumers via a gateway using NDN packet forwarding (the red line shows this data flow). The consumer at the other end can access the content of interest via an IoT application. 
The gateway connects the EDs and NDN to receive EDs' packets and publish the contents of interest to NDN under its namespace \cite{ICNSurvey,NDN_IoT}. An MECD can be a resource-rich, small-scale data center that is co-located with the gateway and is granted the privilege via gateway for executing the costly computing tasks such as the packet signing. 
The gateway can take over the responsibilities of the MECD if it has sufficient computing power for the signing task offloading. As shown in the figure, MECD/gateway can communicate with EDs directly or via a trusted cluster head (CH) node.

At an ED, the NDN-IoT data packet can be encapsulated at the application layer under its own prefix and finally sent down to the physical layer for analog RF signal emission. 
We assume that an ED's imperfect RF front-end can process signals using the in-phase/quadrature-phase (IQ) processing, where the signals suffer from the imbalanced gain and phase-shift between I-channel and Q-channel mainly due to the imperfect local oscillator and mixer \cite{bookIQI}.
We use $\theta$ and $\alpha$ to represent phase-shift mismatch and amplitude mismatch, respectively, where the independent $\theta$ and  $\alpha$ follow the uniform distributions , $\theta\sim U(-\theta_m, \theta_m)$ and $\alpha\sim U(-\alpha_m, \alpha_m)$, and $\theta_m>0$ and $\alpha_m>0$ are regulated by the RF manufacturer/market\cite{PDFIQI,AccessPeng,IQIrangebook}. We consider that all EDs in the MECD's coverage are produced under the same maximum $\alpha$ and $\theta$ requirement. At the receiver side, the received signals with IQI effect can be represented as \cite{IQI_est,AccessPeng}

\begin{align}\label{Rxsignal}
\!\!\!\!\mathbf z&=[c_1 \mathbf H (1+(1+\alpha_{\text{}})e^{j\theta_{\text{}}})+c_2\mathbf H^*(1-(1+\alpha_{\text{}})e^{-j\theta_{\text{}}})] \frac{\mathbf x}{2}\nonumber \\
&+[c_1 \mathbf H (1-(1+\alpha_{\text{}})e^{j\theta_{\text{}}})+c_2\mathbf H^*(1+(1+\alpha_{\text{}})e^{-j\theta_{\text{}}})]\frac{\mathbf x^*}{2},
\end{align}
where $\mathbf x$ denotes the transmitted symbol vector in a time instant, 
$\mathbf H$ is the corresponding channel gain matrix, and $c_1$ and $c_2$ are two constants representing the IQI of the receiver. 
Usually, the IQI estimation is a mandatory function of the signal receiver for IQI compensation, and the estimated IQI parameter can be generally represented by
\begin{align}\label{aEst}
\hat a=a+n_a,
\end{align}where $\hat a$ and $a$ denote the estimated and actual values of the constant IQI parameter, respectively, and $n_a$ denotes the estimation noise that follows $n_a\sim \mathcal N(0, \sigma^2)$ \cite{reportIQI5,imperfection2}. Here, $a$ is a function of $\theta$ and $\alpha$; the different forms of $a$ will be discussed in Section \ref{MEBsection}.

Typically, NDN-IoT requires that all data packets are signed by the private signing key ($PvtK_{\text{ED}_{A}}$) of the resource-constrained $\text{ED}_A$ for data provenance authentication, where $\text{ED}_A$ denotes the end device A. The packet signing is based on asymmetric encryption, which employs, for example, ECC and RSA in most NDN networks. This could impose a heavy computational workload on the constrained $\text{ED}_A$ and significantly delay the packet uploading. It is assumed that EDs cannot afford the per-packet-level asymmetric encryption when preparing the packets due to their limited computation capacities; meanwhile, the MECD is assumed to be unconstrained. The private-public key pair and certificate generation and distribution are beyond the scope of our paper; interested readers can refer to \cite{6843232,NDNsmartcity,zhang2017ndn}. 

This paper focuses on how to replace the task of signing NDN packets at the EDs with the following two steps:
\begin{itemize}
\item Step 1: MECD first authenticates an ED using IQI-based PHY-ID accurately 1) under the dense IoT environment and 2) without imposing excessive computational complexity on this ED.

\item Step 2: MECD offloads the signing task from the authenticated ED through accepting parts of the unsigned packets and completes the packet signing.
\end{itemize}

\subsection{Threat Model and Assumptions}

Given the vulnerability of resource-limited EDs, we consider that the DDoS attacker's objective is to inject poisoned contents into the NDN cache via impersonating the legitimate EDs. We assume that the attackers with high computing power can compromise or even reconstruct the signing key of an ED within the lifetime of signature through key leaking, replaying the captured packets, brute-force computation, etc. 
It is assumed the attackers can be aware of the current pending interest for particular content or can anticipate the popular interests of consumers. 
Hence, such attackers are able to produce poisoned contents under a popular name prefix, sign contents using the compromised key, and try to publish the poisoned data packets on NDN. 
We suppose that the consumers know the trust schema for data provenance verification and that the trust anchors do not change over time (i.e., consumers know how to construct valid
certification chains).
Consequently, although the consumer or NDN router can reach the correct certificate along the trust chain, none of them can determine the signature of attacker as invalid if purely relying on the traditional encryption-based method; thus, they will accept the poisoned contents without triggering any alarm.  In addition, we assume that the EDs' DDoS attacks targeting MECD without using any identity impersonation/spoofing methods will be detected by the MECD.

\subsection{Framework of the Proposed Offline/Online Counter Solution and Assumptions}

To accomplish the aforementioned two-step strategy, we propose to integrate the IQI-based PHY-ID into the existing signature-based NDN-IoT authentication scheme with the aid of MECD, which consists of an offline phase and an online phase. In the following, as MECD and gateway are co-located, we use the two terms interchangeably when presenting the communication procedures for simplicity.

\begin{figure}[!t]
\centering
\includegraphics[width=90mm]{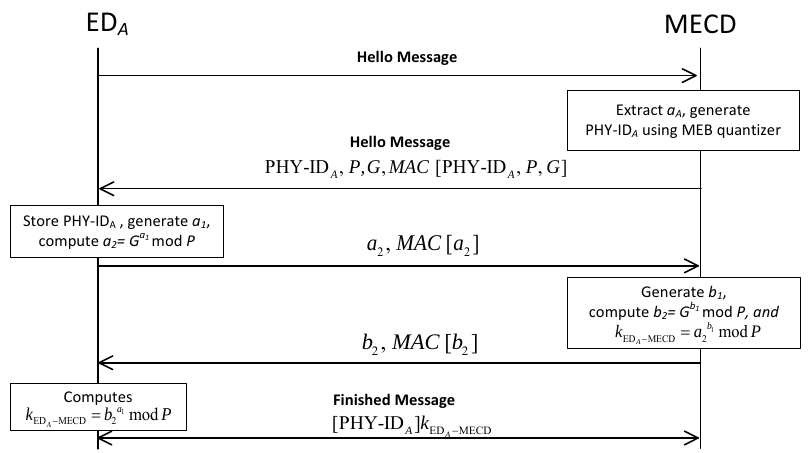}
\caption{Main procedures of the offline phase.}
\label{offlinephase}
\end{figure}

\subsubsection{Offline phase}
The offline phase is responsible for preparing the MEB quantization rule, as well as generating and registering the PHY-ID of the legitimate EDs, as shown in Fig. \ref{offlinephase}. 
We consider to address a legitimate ED$_A$ with IQI parameter $a_A$ as an example.
Since the IQI of $\text{ED}_A$ automatically affects all transmitted RF signals, the MECD is able to extract $a_A$ from any received packets of $\text{ED}_A$ (Hello Message in the figure) or obtain $a_A$ through the relaying of CH. Then, MECD can input $a_A$ into the proposed MEB quantizer to generate $q(\hat a_A)$ and $\text{PHY-ID}_A$ of $\text{ED}_A$ and send $\text{PHY-ID}_A$, two primitive numbers $P$ and $G$, and $MAC[\text{PHY-ID}_A, P, G]$ back to ED$_A$, where $MAC [\cdots]$ denotes the hashed-message authentication code (MAC) and $q(\cdot)$ is a quantization function defined as
\begin{align}\label{quantization} 
q(\hat a)= a_{ [m]}, \ \ \ m=1, 2\cdots M
\end{align}
where $M$ denotes the number of quantization levels. We define $\text{PHY-ID}\triangleq hash(a_{[m]})$, where $hash(\cdot)$ denotes the hash function.
After that, ED$_A$ and MECD need to establish a shared session key ($k_{\text{ED}_A-MECD}$), for example, using Diffie-Hellman key exchange in our case. 
Note that if $k_{\text{ED}_A-MECD}$ is generated using a technique other than Diffie-Hellman key exchange, then $P$ and $G$ do not need to be included in the ``Hello Message''.
ED$_A$ and MECD can exchange $[\text{PHY-ID}_A]_{k_{\text{ED}_A-\text{MECD}}}$ to confirm whether the same $k_{\text{ED}_A-\text{MECD}}$ and PHY-ID$_A$ are shared through the ``Finished Message'', where $[\cdots]_{k}$ denotes the message encrypted by $k$. Finally, MECD registers $(a_A, q(a_A))_{\text{PHY-ID}_A}$ in the whitelist as legitimate and creates a mapping relationship between PHY-ID$_A$ and ED$_A$'s name prefix like ``/ndn/ucla/.../sensor/A/voltage/....'', where /sensor/A/voltage denotes the voltage data of ED$_A$. Thereafter, ED$_A$ needs to add this registered PHY-ID$_A$ to the frame header of its medium access control layer when preparing the data packets if ED$_A$ requires the help of the MECD for signature delegation.

We assume that MECD knows $\theta_m$ and $\alpha_m$ when determining the MEB quantization rule and that all devices are capable of completing the key exchange and enough times of hashed-MAC. The public key of MECD/gateway can be obtained from an authorized manager by any entity who needs to verify its signature.
It is also assumed that all devices are trusted during the offline preparation. The authentication and confidentiality orientated encryptions are used to secure the communication between MECD and CH if a direct link between the ED and MECD does not exist.

\begin{figure}[!t]
\centering
\includegraphics[width=90mm]{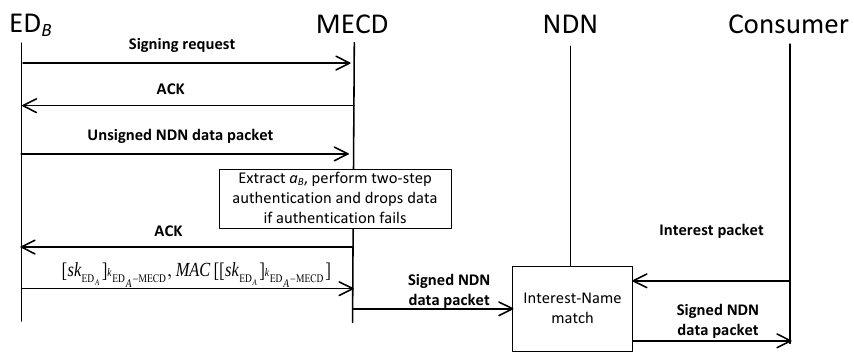}
\caption{Main procedures of the online phase.}
\label{onlinephase}
\end{figure}

\subsubsection{Online phase}
In the online phase, we consider that an IoT $\text{ED}_B$ claiming PHY-ID$_A$ tries to send the unsigned NDN-IoT data packets to the MECD, as shown in Fig. \ref{onlinephase}. After exchanging the signing request and ACK messages, ED$_B$ can verify MECD's signature in the ACK and sends unsigned NDN data packets to MECD. The
MECD can obtain $a_B$, which is physically associated with the RF of the current $\text{ED}_B$ and, thus, can examine the claimed PHY-ID$_A$ using our two-step authentication (presented in Section \ref{twostep}). If authentication fails, the unsigned NDN data packets are rejected. If the identity is verified, MECD can confirm that the unsigned packets are truly sent by the legitimate $\text{ED}_A=\text{ED}_B$, obtain the signing key $SK_{\text{ED}_A}$ in the form of $[SK_{\text{ED}_A}]_{k_{\text{ED}_A-MECD}}, MAC[[SK_{\text{ED}_A}]_{k_{\text{ED}_A-MECD}}]$, execute the costly signing task on behalf of the validated $\text{ED}_B$ and cache the signed data packet in the content store (CS) of an NDN node. At the other end, the NDN router may receive the consumer's interest packet and check CS for an interest-name match. If a match is found, the interest request can be satisfied by feeding the cached contents to the consumer. If one is not found, pending interest table (PIT) lookup could be performed to see if a previous unsatisfied interest with the same name is still pending. Only if a PIT entry does not exist, the interest is forwarded according to the forwarding information base, and a new PIT entry is created.

\subsubsection{Assumptions}
It is assumed that IQI estimation is mandatory at the signal receiver (MECD/CH), for example, using \cite{case1IQIest,case2IQIest,IQI_est,SIoTVTC}.
Therefore, MECD can process the readily available $\hat a$ in both the offline and online phases. 
The noise term $n_a$ in \eqref{aEst} can be ignored in offline registration. The hardware-level $a$ can be stable as a constant during a period that is much longer than a regular authentication session, and thus $n_a$ can be remarkably eliminated based on increasingly many samples of $\hat a$, e.g., using diversity combining techniques. However, this $n_a$ is not negligible in a specific online authentication test due to insufficient samples of this ED's $\hat a$.
All devices cannot be physically captured/broken without triggering any alarm, and the cooperative MECD/CH are always trusted. This study focuses mainly on the perspective of authenticating the data producer ED. Notably, our method can also be applied to the data consumer since the RF parameters can also be extracted from the NDN interest packet if it is transmitted by the wireless node-based consumer. Since signing the interest packet is not mandatory in NDN, the consumer-related authentication issues are not discussed. In addition to the authentication, we suppose that the data integrity and confidentiality are protected by the hash-based encryption and the session key-based encryption, respectively, and therefore, we do not discuss them in this paper. 

In the next sections, we present several key techniques according to the order of process flow, including the MEB-based PHY-ID generation (Section \ref{MEBsection}), the seamless integration of PHY-ID into the NDN-IoT signature (Section \ref{integrationsection}), and the online two-step authentication (Section \ref{twostep}). Finally, we combine them into an algorithm in Section \ref{SectionAlgorithm}.

\section{Offline MEB Quantization Rules}\label{MEBsection}

In this section, we derive the MEB quantization rule for PHY-ID generation. Given the probability density functions (PDFs) of $\theta$ and $\alpha$, the EDs' parameters $a$ are random variables with a specific PDF, $f_{A}(\cdot)$. In the case of nonuniform distribution, the distribution of $a$ can be denser in some quantization intervals, implying a higher probability for different $a$ to appear in these intervals. This could result in deteriorated decision accuracy in testing $q(a_A)$ and $q(a_B)$. Thus, it is important to consider the PDF of $a$ in designing the quantizer. 

Generally, the Shannon entropy of $a_{[m]}$ can be given by
\begin{align}\label{entropyH}
H=-\sum_{m=1}^{M}\tilde f_A(a_{[m]})\log_2\tilde f_A(a_{[m]}),
\end{align}where $\tilde f_A(a_{[m]})$ denotes the probability mass function of $a_{[m]}$. 

\newtheorem{theorem}{Theorem}
\newtheorem{lemma}{Lemma}
\begin{lemma}\label{lemma1}
In the authentication context, the minimization of $\tilde f_A(a_{[m]}), m=1, 2,\cdots, M,$ leads to the maximized distinguishability of $a_{[m]}$ and PHY-IDs.
\end{lemma}

As the measurement of randomness, $\max(H)$ can be interpreted as the maximized randomness of the possible $a_{[m]}/$PHY-IDs. For the purpose of distinguishing EDs, it is expected that different EDs' IQI parameters can be quantized into different PHY-IDs with the maximized probability given $f_{A}(\cdot)$. Since $\tilde f_A(a_{[m]})>0$, $\max(H)$ can be obtained by minimizing all $\tilde f_A(a_{[m]})$ based on \eqref{entropyH}.

\begin{theorem}\label{theorem1}
Based on Lemma \ref{lemma1}, the MEB quantizer can be obtained by making the quantization boundaries $b_m,\  m=0, 1, \cdots, M-1$, satisfy 
\begin{align}\label{MEBoundary}
\left\{
\begin{array}{clc}
&\!\!\!\!\!\!\!\!b_{0}=a_{\text{min}},\\
&\!\!\!\!\!\!\!\!b_{m+1}=F_a^{-1}\left(\frac{1}{M}+F_a(b_{m})\right), 
\end{array}\right.
\end{align}
where  $a\in[a_{\text{min}}, a_{\text{max}}]$ and $F_a(\cdot)$ denotes the antiderivative function of $f_A(\cdot)$.
\end{theorem}
\emph{Proof}: See Appendix A.

However, the challenge of using \eqref{MEBoundary} is that $a$ could have different representations due to the diversity of IQI estimation techniques. Generally, the representation of $a$ can be classified into two cases.
\begin{itemize}
 
\item \textit{Case 1:} Ideally,  $\theta$ and $\alpha$ can be directly estimated, implying $a=\theta$ or $a=\alpha$ \cite{case1IQIest}. 

\item \textit{Case 2:} In most practical cases, $a$ is a function of $\theta$ and $\alpha$, implying $a=f(\theta, \alpha)$  \cite{case2IQIest,IQI_est,SIoTVTC}.
\end{itemize}
Using Theorem \ref{theorem1}, the MECD can prepare the MEB quantization rules offline for the two cases as follows. 

\subsection{Offline MEB Quantization Rule for Case 1 $(a=\theta$ or $\alpha)$}

Since $a$ is either $\theta_{}$ or $\alpha_{}$ in \emph{Case 1}, we obtain $a \sim U(a_{\text{min}}, a_{\text{max}})$, where $-a_{\text{min}}=a_{\text{max}}>0$.
Hence, $a$ is zero mean, variance $\sigma_{a}^2=\frac{a_{\text{max}}^2}{3}$, and $f_{A}(x)=\frac{1}{2a_{\text{max}}}, x\in[-a_{\text{max}}, a_{\text{max}}]$. For uniformly distributed input $a$, only a uniform quantizer can satisfy \eqref{MEBoundary};  
thus, the decision boundary of an interval is
\begin{align}\label{bound1}
b_{m}=\left(\frac{2m}{M-1}-1\right)a_{\text{max}}.
\end{align} 

\subsection{Offline MEB Quantization Rule for Case 2 $\left(a=f(\theta, \alpha)\right)$}\label{case2quan}

Without loss of generality, we express $a$ in this case as
\begin{align}\label{afunction}
a=g_{1}(\theta)g_2(\alpha)+c,
\end{align}
where $g_{1}(\theta)$ is a function of $\theta$, $g_{2}(\alpha)$ is a function of $\alpha$, and $c$ is a constant. We use $g_1$ and $g_2$ in the following for simplicity. 

\begin{theorem}\label{theorem2}
\renewcommand{\arraystretch}{1}
Given the independent random variables $\alpha$ and $\theta$ and $g_{1}\in [g_{\text{min}1} g_{\text{max}1}]$ and $g_{2}\in [g_{\text{min}2}, g_{\text{max}2}]$, an ED's IQI parameter $a$ with the expression of \eqref{afunction} has the PDF $f_A(x)$, which can be derived as follows.

\noindent If $g_{\text{min}1}g_{\text{max}2}<g_{\text{max}1}g_{\text{min}2}$,
\begin{align}\label{PDF1g}
\!\!f_A(x)=\left\{
\begin{array}{clc}
\tau_1(x-c),& g_{\text{min}1}g_{\text{min}2}<g<g_{\text{min}1}g_{\text{max}2}\\
\tau_2(x-c), &g_{\text{min}1}g_{\text{max}2}<g<g_{\text{max}1}g_{\text{min}2}\\
\tau_3(x-c), & g_{\text{max}1}g_{\text{min}2}<g<g_{\text{max}1}g_{\text{max}2}
\end{array}
\right.,
\end{align}
\noindent if $g_{\text{min}1}g_{\text{max}2}>g_{\text{max}1}g_{\text{min}2}$,
\begin{align}\label{PDF3g}
\!\!f_A(x)=\left\{
\begin{array}{clc}
\tau_1(x-c),& g_{\text{min}1}g_{\text{min}2}<g<g_{\text{max}1}g_{\text{min}2}\\
\tau_4(x-c), &g_{\text{max}1}g_{\text{min}2}<g<g_{\text{min}1}g_{\text{max}2}\\
\tau_3(x-c), & g_{\text{min}1}g_{\text{max}2}<g<g_{\text{max}1}g_{\text{max}2}
\end{array}
\right.,
\end{align}
\noindent if $g_{\text{min}1}g_{\text{max}2}=g_{\text{max}1}g_{\text{min}2}$,
\begin{align}\label{PDF2g}
f_A(x)=\left\{
\begin{array}{clc}
\tau_1(x-c),& g_{\text{min}1}g_{\text{min}2}<g<g_{\text{min}1}g_{\text{max}2}\\
\tau_3(x-c), & g_{\text{min}1}g_{\text{max}2}<g<g_{\text{max}1}g_{\text{max}2}
\end{array}
\right.,
\end{align}where 
\begin{align}\label{tau1}
&\tau_i(x)=\int_{\kappa_{b_{(i)}}}^{\kappa_{t_{(i)}}} \frac{1}{x}f_{G_1}(x) f_{G_2}(g/x) \text dx, \,\, i=1,2,3,4
\end{align}where
$f_{G_1}(x)$  and $f_{G_2}(x)$ are the PDFs of $g_1$ and $g_2$, respectively, $\bm \kappa_t=[g/g_{\text{min2}},\, g/g_{\text{min2}},\, g_{\text{max1}},\,g_{\text{max1}}]$, and $\bm \kappa_b=[g_{\text{min1}},\, g/g_{\text{max2}},\, g/g_{\text{max2}},\,g_{\text{min1}}]$.
\end{theorem}
\emph{Proof}: See Appendix B.

Let $F_{\tau_1}$, $F_{\tau_2}$, $F_{\tau_3}$, and $F_{\tau_4}$ represent the antiderivative functions of $\tau_1(x-c)$, $\tau_2(x-c)$, $\tau_3(x-c)$ and $\tau_4(x-c)$, respectively, $b_0=a_{\text{min}}$, $G_1=g_{\text{min}1}g_{\text{max}2}$, $G_2=g_{\text{max}1}g_{\text{min}2}$, and $G_3=g_{\text{max}1}g_{\text{max}2}$. Using Theorem \ref{theorem1}, MECD can obtain the MEB boundaries for eq.~\eqref{PDF1g} as 
\begin{align}\label{boundarycase2_1}
\renewcommand{\arraystretch}{1.3}
b_{m+1}\!=\!\left\{
\begin{array}{clc}
&\!\!\!\!\!\!\!\!\!\! F_{\tau_2}^{-1} \left(\frac{1}{M}-F_{\tau_1}\left(G_1\right) + F_{\tau_1}(b_{m})+F_{\tau_2}\left(G_1\right) \right), \\ 
& \ \ \ \ \ \ \ \, \text{if} \ \  b_m<G_1, \left(F_{\tau_1}\left(G_1\right)-F_{\tau_1}(b_{m})\right)<\frac{1}{M}\\
&\!\!\!\!\!\!\!\!\!\! F_{\tau_1}^{-1} \left( \frac{1}{M}+F_{\tau_1}(b_{m})\right), \\ 
& \ \ \ \ \ \ \ \, \text{if}\ \  b_m<G_1,  \left(F_{\tau_1}\left(G_1\right)-F_{\tau_1}(b_{m})\right)>\frac{1}{M}\\
&\!\!\!\!\!\!\!\!\!\! F_{\tau_3}^{-1} \left(\frac{1}{M}-F_{\tau_2}\left(G_2\right) + F_{\tau_2}(b_{m})+F_{\tau_3}\left(G_2\right) \right), \\ 
&        \text{if}\ \  G_1<b_m<G_2,  \left(F_{\tau_2}\left(G_2\right)-F_{\tau_2}(b_{m})\right)<\frac{1}{M}\\
&\!\!\!\!\!\!\!\!\!\! F_{\tau_2}^{-1} \left( \frac{1}{M}+F_{\tau_2}(b_{m})\right), \\ 
&      \text{if}\ \  G_1<b_m<G_2,  \left(F_{\tau_2}\left(G_2\right)-F_{\tau_2}(b_{m})\right)>\frac{1}{M}\\
&\!\!\!\!\!\!\!\!\!\! F_{\tau_3}^{-1} \left(\frac{1}{M}+F_{\tau_3}\left(b_m\right) \right), \\ 
& \ \ \ \ \ \ \ \ \ \ \ \ \ \ \ \ \ \ \ \ \ \ \    \text{if}\ \  G_2<b_m<G_3,  m<M
\end{array}
\right..
\end{align}

\noindent The MEB boundaries for eq.~\eqref{PDF3g} can be given by
\begin{align}\label{boundarycase2_2}
\renewcommand{\arraystretch}{1.3}
b_{m+1}\!=\!\left\{
\begin{array}{clc}
&\!\!\!\!\!\!\!\!\!\! F_{\tau_4}^{-1} \left(\frac{1}{M}-F_{\tau_1}\left(G_1\right) + F_{\tau_1}(b_{m})+F_{\tau_4}\left(G_1\right) \right), \\ 
& \ \ \ \ \ \ \ \, \text{if} \ \  b_m<G_2, \left(F_{\tau_1}\left(G_1\right)-F_{\tau_1}(b_{m})\right)<\frac{1}{M}\\
&\!\!\!\!\!\!\!\!\!\! F_{\tau_1}^{-1} \left( \frac{1}{M}+F_{\tau_1}(b_{m})\right), \\ 
& \ \ \ \ \ \ \ \, \text{if}\ \  b_m<G_2,  \left(F_{\tau_1}\left(G_1\right)-F_{\tau_1}(b_{m})\right)>\frac{1}{M}\\
&\!\!\!\!\!\!\!\!\!\! F_{\tau_3}^{-1} \left(\frac{1}{M}-F_{\tau_4}\left(G_1\right) + F_{\tau_4}(b_{m})+F_{\tau_3}\left(G_1\right) \right), \\ 
&        \text{if}\ \  G_2<b_m<G_1,  \left(F_{\tau_4}\left(G_1\right)-F_{\tau_4}(b_{m})\right)<\frac{1}{M}\\
&\!\!\!\!\!\!\!\!\!\! F_{\tau_4}^{-1} \left( \frac{1}{M}+F_{\tau_4}(b_{m})\right), \\ 
&      \text{if}\ \  G_2<b_m<G_1,  \left(F_{\tau_4}\left(G_1\right)-F_{\tau_4}(b_{m})\right)>\frac{1}{M}\\
&\!\!\!\!\!\!\!\!\!\! F_{\tau_3}^{-1} \left(\frac{1}{M}+F_{\tau_3}\left(b_m\right) \right), \\ 
& \ \ \ \ \ \ \ \ \ \ \ \ \ \ \ \ \ \ \ \ \ \ \    \text{if}\ \  G_1<b_m<G_3,  m<M
\end{array}
\right..
\end{align}

\noindent The MEB boundaries for eq.~\eqref{PDF2g} can be given by
\begin{align}\label{boundarycase2_3}
\renewcommand{\arraystretch}{1.3}
b_{m+1}\!=\!\left\{
\begin{array}{clc}
&\!\!\!\!\!\!\!\!\!\! F_{\tau_3}^{-1} \left(\frac{1}{M}-F_{\tau_1}\left(G_1\right) + F_{\tau_1}(b_{m})+F_{\tau_3}\left(G_1\right) \right), \\ 
& \ \ \ \ \ \ \ \, \text{if} \ \  b_m<G_1, \left(F_{\tau_1}\left(G_1\right)-F_{\tau_1}(b_{m})\right)<\frac{1}{M}\\
&\!\!\!\!\!\!\!\!\!\! F_{\tau_1}^{-1} \left( \frac{1}{M}+F_{\tau_1}(b_{m})\right), \\ 
& \ \ \ \ \ \ \ \, \text{if}\ \  b_m<G_1,  \left(F_{\tau_1}\left(G_1\right)-F_{\tau_1}(b_{m})\right)>\frac{1}{M}\\
&\!\!\!\!\!\!\!\!\!\! F_{\tau_3}^{-1} \left(\frac{1}{M}+F_{\tau_3}\left(b_m\right) \right), \\ 
& \ \ \ \ \ \ \ \ \ \ \ \ \ \ \ \ \ \ \ \ \ \ \    \text{if}\ \  G_1<b_m<G_3,  m<M
\end{array}
\right..
\end{align}

Based on the derivations above, the MEB quantization rule can be determined using Algorithm \ref{case2algorithm}.

\begin{algorithm}[!h]
\caption{Algorithm for the MEB quantization rule.}
\label{case2algorithm}
\begin{algorithmic}[1]
\small

\State \textbf{Inputs:} 

$\theta_m$, $\alpha_m$, $b_0=a_{\text{min}}$, and the expression of $a$.\vspace{1mm}

\State  Calculate $a\in[a_{\text{min}}, a_{\text{max}}]$ using $\theta_m$ and $\alpha_m$. 

\State Rewrite $a$ to get $g_{1}$, $g_2$, and $c$ using \eqref{afunction}. 

\State  Given that $\theta_{\text{}}\sim U(-\theta_m, \theta_m)$ and $\alpha_{\text{}}\sim U(-\alpha_m, \alpha_m)$, $f_{G_1}(x)$ of $g_{1}$ and $f_{G_2}(x)$ of $g_2$ can be derived, respectively.

\If{$g_{\text{min}1}g_{\text{max}2}<g_{\text{max}1}g_{\text{min}2}$}
\State  Obtain $f_A(x)$ using  \eqref{PDF1g}. 
\For{$m \gets 0$ to $M-1$}
\State Obtain $b_{m+1}$ according to \eqref{boundarycase2_1}.
\EndFor
\ElsIf{$g_{\text{min}1}g_{\text{max}2}>g_{\text{max}1}g_{\text{min}2}$}
\State Obtain $f_A(x)$ using  \eqref{PDF3g}.
\For{$m \gets 0$ to $M-1$}
\State Obtain $b_{m+1}$ according to \eqref{boundarycase2_2}.
\EndFor
\ElsIf{$g_{\text{min}1}g_{\text{max}2}=g_{\text{max}1}g_{\text{min}2}$}
\State Obtain $f_A(x)$ using  \eqref{PDF2g}.
\For{$m \gets 0$ to $M-1$}
\State Obtain $b_{m+1}$ according to \eqref{boundarycase2_3}.
\EndFor
\EndIf

\State  \textbf{Outputs:} $b_{m+1}$
\end{algorithmic}
\end{algorithm} 

\begin{figure*}[!htb]
\centering
\includegraphics[width=160mm]{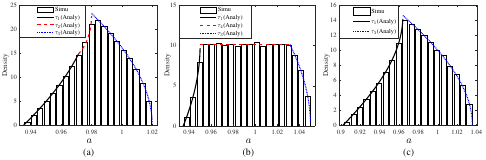}
\caption{Simulated density and analytical density of $a$: (a) $\theta_{m}=\frac{5\pi}{36}$, $\alpha_{m}=0.04$; (b) $\theta_{m}=\frac{\pi}{12}$, $\alpha_{m}=0.1$; (c) $\theta_{m}=\frac{\pi}{6}$, $\alpha_{m}=0.0718$.}
\label{density}
\end{figure*}

\emph{\underline{Case study:}} We present a case study to show how to practically apply our general Algorithm 1.
We consider $a=\frac{1}{2}+\frac{1}{2}(1+\alpha)\cos\theta$ since this representative $a$ can be available in most IQI estimation methods. For example, study \cite{case2IQIest} can produce $\eta_t=\frac{1-u_{\text{}}}{u_{\text{}}}$, where $\Re(u)=a$. Studies \cite{IQI_est} and \cite{SIoTVTC} can produce $k_1, k_2, k_3$, and $k_4$, and the same $\eta_t$ can be obtained by $\eta_t=\frac{k_3}{k_1}$. Through some manipulations, we can obtain $a=\frac{1}{\Re(\eta_t+1)}$.

Fig.~\ref{density} compares the simulation and analytical density of this $a$, where (a), (b), and (c) can correspond to eq. \eqref{PDF1g}, \eqref{PDF3g}, and \eqref{PDF2g}, respectively.  
Regarding the analytical results, we first make $g_{1}=\frac{1}{2}(1+\alpha)$, $g_2=\cos\theta$, $c=\frac{1}{2}$ using \eqref{afunction} so that $f_{G_1}(x)$ and $f_{G_2}(x)$ can be given by
\begin{align}\label{PDFg1g2}
&f_{G_1}(x)=\frac{1}{\theta_m\sqrt{1-x^2}}, \ \ x\in[\cos\theta_m, 1]\nonumber \\
&f_{G_2}(x)=\frac{1}{\alpha_m}, \ \ x\in\left[\frac{1-\alpha_m}{2}, \frac{1+\alpha_m}{2}\right].
\end{align}
Substituting \eqref{PDFg1g2} into \eqref{PDF1g}, \eqref{PDF3g}, and \eqref{PDF2g}, $f_A(x)$ of the three cases can be obtained by comparing $(1+\alpha_m)\cos\theta_m$ and $1-\alpha_m$, and 
$\tau_1(x)$, $\tau_2(x)$, $\tau_3(x)$, and $\tau_4(x)$ become
\begin{align}\label{func1}
&\tau_1(x)=\frac{1}{\theta_m\alpha_m}\ln\left(\frac{ \frac{x}{\cos\theta_m}+\sqrt{\left(\frac{x}{\cos\theta_m}\right)^2-x^2}}{\frac{1-\alpha_m}{2}+\sqrt{\left(\frac{1-\alpha_m}{2} \right)^2-x^2}}\right) \\\label{func2}
&\tau_2(x)=\frac{1}{\theta_m\alpha_m} \ln\left(\frac{\frac{1+\alpha_m}{2} +\sqrt{\left(\frac{1+\alpha_m}{2} \right)^2-x^2}}{\frac{1-\alpha_m}{2} +\sqrt{\left(\frac{1-\alpha_m}{2} \right)^2-x^2}}\right)
\end{align}
\begin{align}\label{func3}
&\tau_3(x)=\frac{1}{\theta_m\alpha_m} \ln\left(\frac{1+\alpha_m}{2x} +\sqrt{\left(\frac{1+\alpha_m}{2x}\right)^2-1}\right) \\\label{func4}
&\tau_4(x)=\frac{1}{\theta_m\alpha_m} \ln\left(\frac{1}{\cos\theta_m} +\sqrt{\left(\frac{1}{\cos\theta_m}\right)^2-1}\right).
\end{align}
The segmented functions $\tau_1$, $\tau_2$, $\tau_3$, and $\tau_4$ represent the analytical results in Fig.\ref{density}. Regarding the simulation results, we randomly generate $10^5$ samples of $\theta$ and $\alpha$ according to their PDFs, and plot  the corresponding PDF of $a$ statistically. As expected, the analytical and simulation results closely match, which validates our derivations in Theorem \ref{theorem2}.

We take the case of $(1+\alpha_m)\cos\theta_m<1-\alpha_m$ as an example. In practical offline preparations, MECD can use Algorithm 1 to obtain $g_{1}(\alpha)=\frac{1}{2}(1+\alpha)$, $g_2(\theta)=\cos\theta$, $c=\frac{1}{2}$ using \eqref{afunction}, calculate $f_A(x)$ using \eqref{PDF1g} \eqref{func1} \eqref{func2} \eqref{func3}, and finally determine the MEB boundaries by substituting $f_A(x)$ into \eqref{boundarycase2_1}.

\section{Integration of PHY-ID Into The NDN-IoT Signature Scheme}\label{integrationsection}

\begin{figure}[!t]
\centering
\includegraphics[width=90mm]{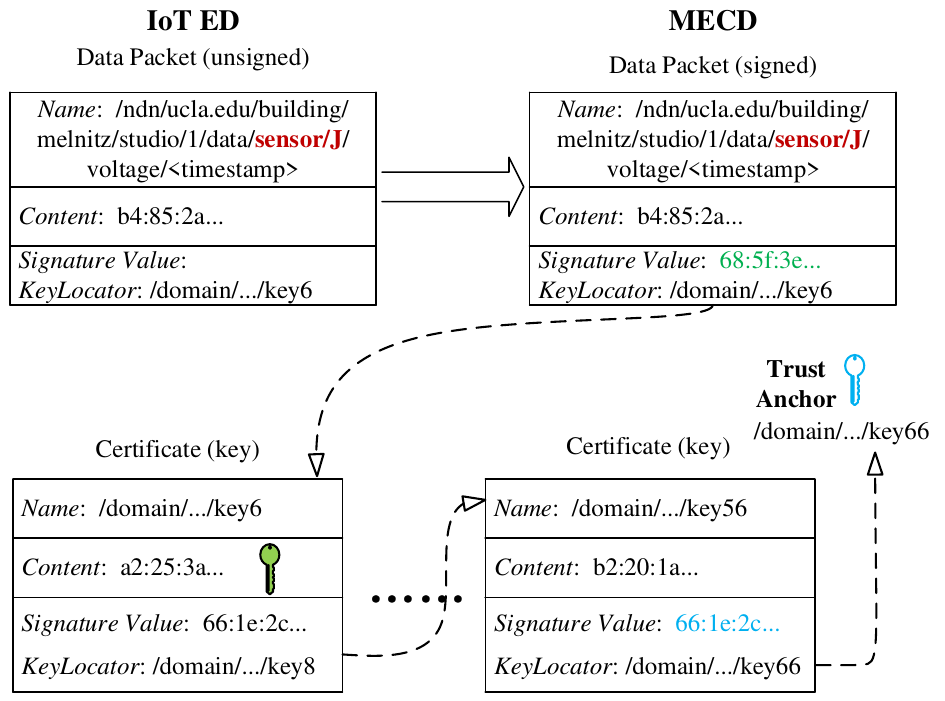}
\caption{Proposed NDN-IoT authentication.}
\label{PHY-NDN}
\end{figure}

In this section, we consider the offline preparation has been completed, and ED$_A$'s PHY-ID$_A$ has been generated using our MEB quantization and registered by MECD. 
Fig. \ref{PHY-NDN} illustrates our NDN signature scheme. 
Both the NDN data packet and certificate packet have the same format, which consists of the \emph{Name, Content, Signature Value} and \emph{KeyLocator} fields. 
The primary purpose of \emph{Name} is to facilitate the name-indexed content identification and routing. 
To this end, application-driven NDN can define the hierarchical structure \emph{Name}, e.g., /ndn/ucla.edu/building/melnitz/studio/1/data/sensor/J/voltage
/$<$timestamp$>$\cite{6843232}, where the segment sensor/J can denote ED$_A$ at the network layer.
The \emph{Signature Value} is the result of signing on the \emph{Content}, which requires asymmetric encryption. The name of the signing key is put in \emph{KeyLocator} so that the verifier can retrieve the public key in the same manner as it can retrieve the signed data packet for signature verification.  

We consider that $\text{ED}_B$ is claiming PHY-ID$_A$ in the medium access control layer frame header of the unsigned NDN data packets to the MECD. As shown in the figure, the NDN data packet leaves \emph{Signature Value} empty. 
Since the actual IQI of $\text{ED}_B$ inevitably affects all unsigned packets at physical layer, the MECD is able to obtain the ED$_B$-specific $a_{B}$ and generate $\mathcal B=(a_B, q(a_B))_{\text{PHY-ID}_B}$. 
MECD can extract the claimed $\text{PHY-ID}_A$ and check if it can be mapped to the name ``...sensor/J...'' that falls in its name prefix. If correct, MECD can use $\text{PHY-ID}_A$ as an index to find the registered $\mathcal A=(a_A, q(a_A))_{\text{PHY-ID}_A}$, and the identity of $\text{ED}_B$ can be verified by comparing $\mathcal A$ and $\mathcal B$ using our two-step authentication method.
In brief, if $q(a_A)\neq q(a_B)$, it means the claimed $\text{PHY-ID}_A$ does not belong to this ED$_B$, and thus MECD should reject the NDN data packet of ED$_B$. If $q(a_A)= q(a_B)$, MECD will further compare $a_A$ and $a_B$. If they are identical, MECD can start to sign on every data packet on behalf of the validated ED$_A$. Otherwise, MECD rejects ED$_B$. In NDN, the name in \emph{KeyLocator} of the signed data packet could point to a certificate for retrieving the public key. The certificate can be linked to the next one until it finally reaches the pre-agreed trust anchor. The multiple retrieved keys between the unsigned data packet and the trust anchor form a trust chain to protect the authentication scheme \cite{thesisNDN}.

\section{Online Two-Step Authentication}\label{twostep}

This section presents the accurate online two-step authentication technique in the scenario that a large number of EDs are densely deployed. We consider that an ED$_B$ ($\text{ED}_B\neq\text{ED}_A$) moves to the coverage of the MECD and claims PHY-ID$_A$. As aforementioned, MECD can obtain $(\hat a_B, q(\hat a_B))_{\text{PHY-ID}_A}$ from ED$_B$'s packets and find $(a_A, q(a_A))_{\text{PHY-ID}_A}$ from the whitelist.

In step 1, we directly compare $q(\hat a_B)$ with the registered $q(a_A)$. Since our MEB quantization minimizes the probability of two IQI parameters falling into the same quantization interval, $q(a_A)\neq q(\hat a_B)$ can be produced in most cases.
However, if $a_A$ and $a_B$ are very close, it is possible to obtain $q(a_A)=q(\hat a_B)$. The chance of encountering this problem can increase when a large number of EDs are involved. To resolve this problem, we further propose another virtual boundary, $b_v$, for the step 2 examination.  

In step 2, this problem can be mathematically formulated as $a_A\neq a_B$, but $q(a_A)$ and $q(a_B)$ fall in the interval $[b_m, b_{m+1}]$. We assume $a_A$ and $a_B$ have been registered so that $\text{MECD}$ knows $a_A$ and $a_B$. Note that the assumption of known $a_B$ can be relaxed without changing the authentication performance, as will be discussed later.
For presentation simplicity, we set that ${\hat{a}}_x={a}_x+{n_a}$ is under the step 2 test, where $x=A$ or $B$. 
Since the same $a_x$ affects all RF signals or even all subcarriers of an OFDM signal \cite{IQIAFproof,sameIQIaffect}, it is considered that $N_s$-independent estimation samples of $\hat a_x$ can be conveniently observed from $N_p$ packets. At the MECD, the offset between $a_A$ and the $k$-th $\hat a_{x[k]}$ is 
\begin{align}\label{offset}
y_{[k]}=\hat a_{x[k]}-a_A=a_{\Delta}+n_{a[k]}, \,\, k=1,2,\cdots,N_s
\end{align}where $a_{\Delta}=a_x-a_A$. 
Based on \eqref{offset}, a binary hypothesis testing can be modeled by
\renewcommand{\arraystretch}{1}
\begin{align}\label{binaryHypo}
\left\{
\begin{array}{clc}
\mathcal{H}_0: & a_{\Delta}= 0 \\
\mathcal{H}_1: & a_{\Delta} \neq 0
\end{array}
\right.,
\end{align}where $\mathcal{H}_0$ implies $a_x=a_A$, and $\mathcal{H}_1$ means $a_x=a_B$. Therefore, $y_{[k]}\sim \mathcal N(0, \sigma^2)$ under $\mathcal{H}_0$ with the likelihood function $p_Y( y|\mathcal{H}_0)$ and $y_{[k]}\sim\mathcal N(a_{\Delta}, \sigma^2)$ under $\mathcal{H}_1$ with the likelihood function $p_Y( y|\mathcal{H}_1)$.
Since $q(a_A)=q(\hat a_B)$, $a_A$ is extremely close to $a_B$, implying the difficulty in detecting the nonzero $a_{\Delta}$ under $\mathcal{H}_1$.

\subsubsection{Known $\sigma^2$}
If $\sigma^2$ is known, the Neyman-Pearson lemma can be used to differentiate the hypotheses since it can maximize the probability of detecting $\mathcal{H}_1$ given a false alarm rate \cite{Kay}. The likelihood ratio test (LRT) based on $N_s$ samples is $
\left(\frac{p_Y( y|\mathcal{H}_1)}{p_Y(y|\mathcal{H}_0)}\right)^{N_s}$, which produces  
\begin{align}\label{LRTresult}
L_{\text{NP}}=\frac{r\sum_{k=1}^{N_s}y_{[k]}}{a_{\Delta}N_s} \mathop{\lessgtr}\limits_{\mathcal{H}_1}^{\mathcal{H}_0} b_{v},
\end{align}
where $r=\frac{a_{\Delta}^2}{\sigma^2}$ is the offset-to-noise ratio, $b_{v}$ denotes the virtual boundary for differentiating $a_A$ and $a_B$, and $L_{\text{NP}}\sim\mathcal N\left(r, \frac{r^2\sigma^2}{N_sa_{\Delta}^2}\right)$ under $\mathcal{H}_1$. If $L_{\text{NP}}>b_{v}$, $a_A$ and $a_B$ (i.e., $\mathcal H_1$) can be differentiated; otherwise, $\mathcal H_0$ is determined. Given a  required false alarm rate $\rho$, the $b_v$ can be given by
\begin{align}\label{threshold}
b_v=Q^{-1}(\rho)\sqrt{\frac{r}{N_s}},
\end{align}where $Q^{-1}(\cdot)$ is the inverse of the Q-function, and $\rho$ can be mathematically defined by
\begin{align}\label{pfa}
\rho\!=\!\!\int_{b_v}^{\infty}\!\sqrt{\frac{N_s}{2r\pi}}\exp\left(-\frac{N_sx^2}{2r}\right) \text d x\!.
\end{align}
The differentiation rate is defined by 
\begin{align}\label{DiffRate}
r_{\text{D}}=Pr(L_{\text{NP}}>b_v|\mathcal{H}_1)=Q\left(\sqrt{rN_s}\left(\frac{b_v}{r}-1\right)\right).
\end{align}
More generally, if $a_B$ is not registered, $a_{\Delta}$ becomes unknown. In this case,  \eqref{LRTresult} can become $L_{\text{NP}}=\frac{1}{N_s}\sum_{k=1}^{N_s}y_{[k]}$ and \eqref{threshold} can become $b_v=\frac{\sigma}{\sqrt{N_s}}Q^{-1}(\rho)$. As a result, the differentiation performance remains unchanged since the new $L_{\text{NP}}$ and $b_v$ can result in the same $r_{\text D}$ in \eqref{DiffRate}. 
  
\subsubsection{Unknown $\sigma^2$}  
In practice, the variance of the estimation noise is not always available. If $\sigma^2$ is unknown, we can use the generalized LRT (GLRT) to process the offset samples $y_{[k]}$.
In this case, the GLRT-based hypothesis decision can be made by
\begin{align}\label{GLRT}
L_{\text{GLRT}}=\frac{\left(N_s-1\right)\left(\sum_{k=1}^{N_s}y_{[k]}\right)^2}{N_s\sum_{k=1}^{N_s}(y_{[k]}-\frac{1}{N_s}\sum_{k=1}^{N_s}y_{[k]})^2} \mathop{\lessgtr}\limits_{\mathcal{H}_1}^{\mathcal{H}_0} b_{v},
\end{align}where the PDF of $L_{\text{GLRT}}$ is 
\begin{align}\label{Fdist1}
\!\!\!\!L_{\text{GLRT}}\sim\left\{
\begin{array}{clc}
&\!\!\!\!\!\!\!\!\!\!F(1,N_s-1) \ \ \ \ \ \ \ \ \ \ \, \text{under}\, \mathcal H_0\\
&\!\!\!\!\!\!\!\!\!\!F'(1,N_s-1,\frac{N_sa^2_{\Delta}}{\sigma^2}) \ \ \text{under}\, \mathcal H_1
\end{array}
\right.,
\end{align}where $F$ and $F'$ denote central/non-central Fisher–Snedecor distributions ($F$-distribution). Please see Appendix C for the proofs of \eqref{GLRT} and \eqref{Fdist1}.

Let $F_{L_{\text{GLRT}}}(x|\mathcal H_0/\mathcal H_1)$ denote the cumulative distribution function of $L_{\text{GLRT}}$, the required $\rho$ equals 
\begin{align}\label{pfaGLRT} 
\rho&\!=\!\!1\!\!-\!\!F_{L_{\text{GLRT}}}(b_v, 1, N_s-1|\mathcal H_0)=1-I_{\frac{b_v}{b_v+N_s-1}}\left(\frac{1}{2},\frac{N_s-1}{2}\right)\nonumber \\
&=1-N_s(N_s-1)\int_{0}^{\frac{b_v}{b_v+N_s-1}}(1-t)^{N_s-2}\text dt,
\end{align}where $I_x(d_1,d_2)$ is the regularized incomplete beta function. Given that $d_1$ and $d_2$ are positive integers in our case, $I_x(d_1,d_2)$ can be calculated by $$I_x(d_1,d_2)=\frac{(d_1+d_2-1)!\int_{0}^{x}t^{d_1-1}(1-t)^{d_2-1}\text dt}{(d_1-1)!(d_2-1)!}.$$
Therefore, $b_v$ can be obtained by solving \eqref{pfaGLRT}. Consequently, both $L_{\text{GLRT}}$ and $b_v$ are independent of the unknown $\sigma^2$. 

$r_{\text D}$ can be mathematically computed by
\begin{align}\label{rdGLRT}
&r_{\text{D}}=Pr(L_{\text{GLRT}}>b_v|\mathcal{H}_1)\nonumber \\
&=\!1\!\!-\!\!\sum_{i=0}^{\infty}\!\!\left(\!\!\frac{e^{-\frac{N_sa_{\Delta}^2}{2\sigma^2}}}{i!}\left(\frac{N_sa_{\Delta}^2}{2\sigma^2}\right)^{\!\!i}\right) \!\!I_{\frac{b_v}{b_v+N_s-1}}\!\!\left(\frac{1}{2}\!\!+\!\!i,\frac{N_s-1}{2}\!\right) \\\label{rdapprox}
&\approx1-e^{-\frac{N_s(N_s-1)a_{\Delta}^2}{2\sigma^2(b_v+N_s-1)}}\sum_{i=0}^{N_s-2}\left(\frac{N_s(N_s-1)a_{\Delta}^2}{2\sigma^2(b_v+N_s-1)}\right)^i\frac{1}{i!}\nonumber \\
&\ \ \times \left(\frac{b_v}{b_v+N_s-1}\right)^{i+1}\bigg(1+\sum_{n=1}^{N_s-2-i}\Big(\prod_{m=1}^n\frac{m+i}{m}\Big)\nonumber \\
&\ \ \times \Big(\frac{N_s-1}{b_v+N_s-1}\Big)^n\bigg).
\end{align}Since \eqref{rdGLRT} cannot be used directly for numerical computations, we give its approximation in \eqref{rdapprox} according to \cite{Baharev2017}.

Once $a_A$ and $a_B$ are successfully differentiated using either \eqref{LRTresult} or \eqref{GLRT}, the MECD can adjust $\text{PHY-ID}_A$ and $\text{PHY-ID}_B$. For example, $\text{PHY-ID}_A$ remains unchanged, whereas  $\text{PHY-ID}_B=hash(q(a_B)+\epsilon)$, where $\epsilon$ is a user-defined number for guaranteeing $\text{PHY-ID}_A  \neq \text{PHY-ID}_B$. In addition, one can impose a new boundary $\widetilde b=\frac{a_A+a_B}{2}$ to divide $a_A$ and $a_B$ into two different intervals as $a_A\in[b_m, \widetilde b]$ and $a_B\in[\widetilde b, b_{m+1}]$. In doing so, if $a_B$ appears again, the step 2 test can be avoided since $a_B$ can be differentiated from $a_A$ by the step 1 test. 
Additionally, the $H$ in bits after the insertion of $\widetilde b$ becomes
\begin{align}\label{Hsubboundary}
H=\log_2 M^{\left(1-\frac{1}{M}\right)}C^\frac{1}{C}D^\frac{1}{D}
\end{align}where $C=\frac{1}{\int_{b_{m}}^{\frac{a_B+a_A}{2}}f_A(x)\text dx}$ and $D=\frac{1}{
\int_{\frac{a_A+a_B}{2}}^{b_{m+1}}f_A(x)\text dx}$.

\section{Signing Task Offloading Algorithm}\label{SectionAlgorithm}

It is considered that $\text{ED}_B$ has been authenticated as $\text{ED}_A$ using the method presented in the previous section. Then, the MECD can start to offload the signing task from the verified ED$_A$.

The signing task can be summarized as follows: 
$\text{ED}_A$ has $N_p$ unsigned NDN data packets that need to be signed and published to NDN. 
MECD can consider its own available time to decide to undertake $N_{p1}$ out of the $N_p$ packets and leave $N_{p2}=N_p-N_{p1}$ packets to ED$_A$.
We assume that the computational workload of signing each packet is $B_s$ (in the number of bits), and the signing of all $N_p$ packets should be completed before the deadline $T$. However, the MECD is considered to be available only for the duration $\phi T$, where $0<\phi<1$. $f_{\text{MEC}}$ and $f_{\text{ED}}$ are the fixed CPU-cycle frequency per second of MECD and $\text{ED}_A$, respectively, and $C_b$ is the number of CPU cycles required for computing 1 bit of the input task. Since all signed NDN data packets need to be published to NDN through gateway/MECD regardless of whether the offloading is performed or not, there is no need to consider the time cost for additional task uploading/downloading between the MECD and ED$_A$. The time difference between ED$_A$ and MECD for completing the signing of the same $N_P$ packets is given by
\begin{align}\label{timeReduce}
&\Delta t=N_pB_sC_b\left(\frac{1}{f_{\text{ED}}}-\frac{1}{f_{\text{MEC}}}\right).
\end{align}

Recall the online phase procedure (Fig.\ref{onlinephase}). ED$_A$ sends the signing request to MECD. Then, MECD considers the partitioning of the signing task in the packet granularity and replies that the first $N_{p1}$ unsigned packets can be sent.
It is assumed that the MECD can complete the signing encryption of $N_{p1}$ packets in time $t_1$ and that ED$_A$ can complete up to $N_{p2}$ packets within $t_2$.  
This procedure can be mathematically summarized as a minimax problem as
\begin{align}\label{partition}
&\ \ \ \ \ \ \ \ \min_{N_{p1}, N_{p2}} \max\left(  t_1, t_2\right)\\\label{t1constrain}
&\text{s.t.} \ \ \   0\leqslant t_1=\frac{N_{p1}B_sC_b}{f_{\text{MEC}}}\leqslant\phi T, \\\label{t2constrain}
&\ \ \ \ \ \   0\leqslant t_2=\frac{N_{p2}B_sC_b}{f_{\text{ED}}}\leqslant T, \\\label{constrain3}
&\ \ \ \ \ \ \ \ \ N_{p1}+N_{p2}= N_P.
\end{align} 
Ideally, setting $t_1= t_2$ can achieve the shortest time. If the calculated $N_{p1}$ and $N_{p2}$ are not integers,  we intentionally use $\lceil N_{p1} \rceil$ and $\lfloor N_{p2} \rfloor$ since $\frac{B_sC_b}{f_{\text{ED}}}>\frac{B_sC_b}{f_{\text{MEC}}}$. 

Although the MECD can help complete the signing task efficiently, it may introduce the single-point-of-failure problem into the system as the MECD may become out of service. To avoid this, the use of MECD for task offloading is optional in our system, and the MECD will accept ED's signing request only when the MECD is available, as shown in the online procedures.

Finally, the whole signing task offloading procedures with our offline MEB preparations and online authentication are summarized in Algorithm \ref{MEBalgorithm}.

\begin{algorithm}[!h]
\caption{Algorithm of offline/online phases and signing task offloading.}
\label{MEBalgorithm}
\begin{algorithmic}[1]
\small

\State \noindent\textbf{\underline{\emph{Offline Phase}}:}
\State MECD determines $b_m$ using \eqref{bound1} for \emph{Case 1} or using Algorithm  \ref{case2algorithm} for \emph{Case 2}.

\State MECD registers the legitimate ED$_A$ and defines $\mathcal A=(a_A, q(a_A))_{\text{PHY-ID}_A}$, where $q(a_A)\in\mathbb B_A=[b_m,\, b_{m+1}]$.

\State  \noindent\textbf{\underline{\emph{Online Phase}}:}
\State Quantize $\hat a_B$ to obtain $q(\hat a_B)\in\mathbb B_B$.

\If{$\mathbb B_A\neq\mathbb B_B$}
\State MECD rejects ED$_B$ in the step 1 test.
\Else
\State Compute offset $y_{[k]}$ using \eqref{offset} for step 2 test.

\If{$\sigma^2$ is known }

\State Obtain virtual boundary $b_v$ using \eqref{threshold} and test the ED$_B$ using \eqref{LRTresult}.

\ElsIf{$\sigma^2$ is unknown}
\State Obtain $b_v$ by solving  \eqref{pfaGLRT} and test ED$_B$ using \eqref{GLRT}.
\EndIf 
\If{Step 2 test is not passed} 
\State MECD rejects ED$_B$.
\Else
\State \!\!\!\!\!\!\!\!\!\!\!\!\!\!\!\!\!\!\noindent\textbf{\underline{\emph{Signing Task Offloading}}:}
\State MECD accepts ED$_B$'s unsigned NDN packets for signing task offloading using \eqref{partition}, \eqref{t1constrain}, \eqref{t2constrain}, and \eqref{constrain3}. 

\EndIf
\EndIf
\State End of the algorithm.
\end{algorithmic}
\end{algorithm}

\section{Performance Evaluation}\label{evaluationsection}

In this system evaluation section, we first show the performance  improvements in terms of the authentication accuracy and authentication time efficiency. Then, we analyze the robustness performance in defending against several malicious impersonation attacks.

\subsection{Numerical Results}

We consider the OFDM communication between IoT EDs and MECD. The QPSK modulated signal with $512$ sub-carriers is used when preparing every NDN data packet. The transmitted signal is affected by the ED-specific IQI parameters, passes through the multi-path channel, and finally arrives at the MECD, as shown in eq. \eqref{Rxsignal}. We consider that the circulant channel matrix $\mathbf H$ with the
first column formed by an L × 1 channel impulse response vector 
$\mathbf h=[h_{(1)}, h_{(2)}, \cdots, h_{(L)}]^T$, where the length $L=8$. The elements of $\mathbf h$ are i.i.d. complex circularly symmetric Gaussian random variables as $\mathbf h\sim \mathcal{CN}(\mathbf0,2\mathbf I_L)$. Regarding the IQI parameters, $\theta\sim U(-\theta_m, \theta_m)$, $\alpha\sim U(-\alpha_m, \alpha_m)$, where $\theta_m=\frac{5\pi}{36}$, $\alpha_m=0.04$ if not otherwise specified. 
We consider $a=\frac{1}{2}+\frac{1}{2}(1+\alpha)\cos\theta$ as used earlier in the example study for the offline quantization rule preparation and use \cite{AccessPeng} to obtain the estimated $\hat a$ in the online test phase; the other variables, such as $N_p$, $N_s$, $M$, and $r$, will be given later in each specific evaluation figure.

\begin{figure}[!t]
\centering
\includegraphics[width=77mm]{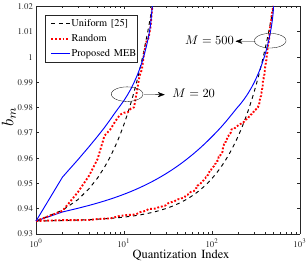}
\caption{MEB quantization boundaries under $M=20$ and $M=500$.}
\label{boundary}
\end{figure}
Fig. \ref{boundary} shows the offline MEB quantization boundaries derived by Algorithm 1 with the comparison of the two other methods, which are named uniform and random. To be specific, the widths of all $M$ quantization intervals are the same regardless of the PDF of $a$ in the uniform method \cite{AccessPeng}. In the random method, all boundaries are randomly picked within $[a_{\text{min}}, a_{\text{max}}]$. It is observed that the MEB boundary value grows faster than those of the other two compared methods. This result occurs because the density of $a$ at the beginning is low, as shown in Fig. \ref{density} (a), resulting in wider intervals for determining the first several $b_m$. However, the width of our MEB quantization intervals narrows as the density increases.
\begin{figure}[!t]
\centering
\includegraphics[width=75mm]{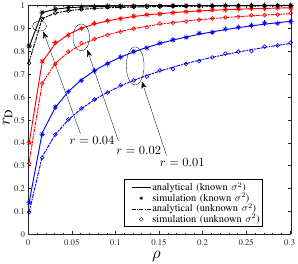}
\caption{Comparison of analytical and simulation results for $r_{\text D}$ vs. $\rho$ under $N_s=400$, $r=0.01, 0.02$ and $0.04$.}
\label{DiffRatefig}
\end{figure}
\begin{figure}[!t]
\centering
\includegraphics[width=80mm]{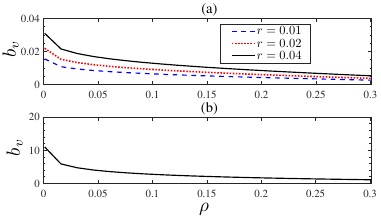}
\caption{Online virtual boundaries: (a) known $\sigma^2$; (b) unknown $\sigma^2$.}
\label{VB}
\end{figure}

The $r_{\text D}$ and the corresponding virtual boundary $b_v$ of our online step 2 test are evaluated in Fig. \ref{DiffRatefig} and Fig. \ref{VB}, respectively. In this simulation, we choose two close IQI parameters, $a_A=1.00166$ and $a_B=1.00177$, indicating that $a_{\Delta}=a_A-a_B$ is equal to only $0.011\%\times a_A$. It is challenging to handle this small $a_{\Delta}$ by the step 1 test since this likely results in $q(a_A)=q(a_B)$, and thus, the step 2 test is needed. It shows that $r_{\text D}$ of both known/unknown $\sigma^2$ cases keep growing with increasing $r$ and $\rho$. This increase occurs because for a fixed $r$, a larger $\rho$ could lead to a smaller $b_v$, as shown in Fig. \ref{VB} (a) and (b). However, $L_{\text{NP}}$ is monotonically increasing with respect to $r$ based on eq. \eqref{LRTresult}, and $L_{\text{GLRT}}$ is independent of $r$ based on \eqref{GLRT}. Fig.\ref{VB} also shows that $b_v$ of case (b) is independent of $r$ compared to $b_v$ of case (a) since $r$ becomes the unknown in case (b).

\begin{figure}[!t]
\centering
\includegraphics[width=77.5mm]{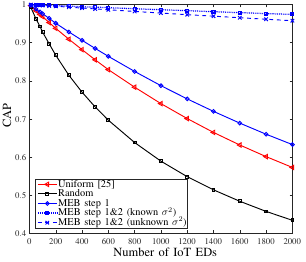}
\caption{Correct authentication probability under different number of IoT end devices, $M=2000$, $\rho=0.01$, $N_s=512$ and $r=0.03$.}
\label{AuthImprovement}
\end{figure}

\begin{figure*}[!t]
\centering
\includegraphics[width=170mm]{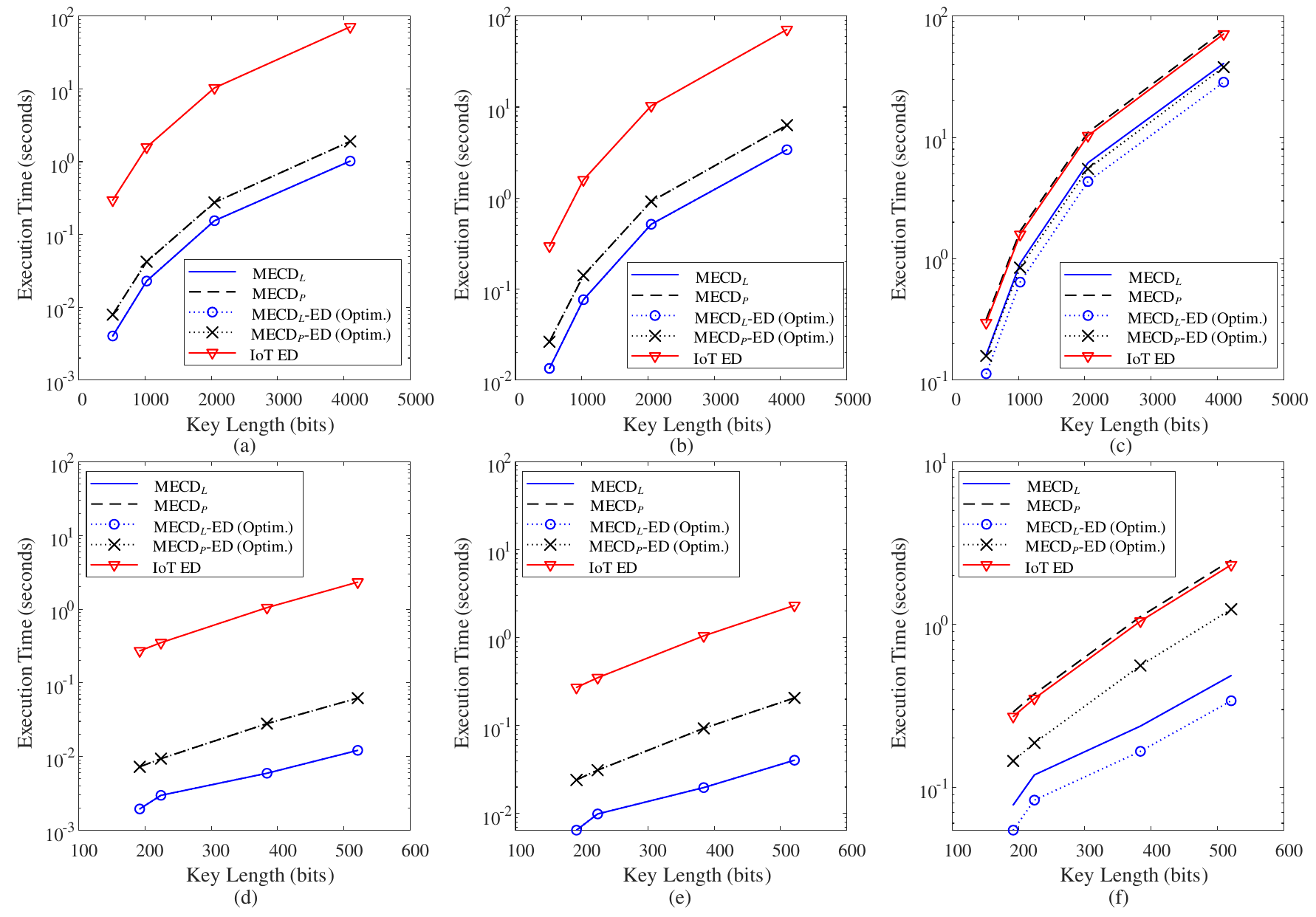}
\caption{The execution time of signing $N_p=10$ NDN packets under RSA and ECC: (a) RSA $\phi=1$; (b) RSA $\phi=0.3$; (c) RSA $\phi=0.025$; (d) ECC $\phi=1$; (e) ECC $\phi=0.3$; (f) ECC $\phi=0.025$.}
\label{RSAtime}
\end{figure*}

After the separate evaluations of the proposed MEB quantization and the online test above, we combine them to evaluate the authentication accuracy performance by executing Algorithm \ref{MEBalgorithm}. To thoroughly test the accuracy in the large number of IoT EDs, we randomly create up to 2000 IoT EDs with different IQI values and register them as the legitimate devices in the offline phase. In each round of the online authentication test, we either randomly generate an attacker or pick up a legitimate one from 2000 IoT EDs. It is further considered that the attacker is sufficiently sophisticated to claim a legitimate PHY-ID and that the attacker's IQI parameter, $a_{ATTK}$, is different from any of the legitimate ones. After the examination of all EDs one by one, we show the correct authentication probability (CAP) in Fig. \ref{AuthImprovement}, where the CAP is defined as the number of correct authentication decisions made by Algorithm \ref{MEBalgorithm} divided by the total authentication attempts. In the step 1 test, CAP of our MEB method outperforms those of the uniform and random methods, which demonstrates the accuracy improvement gained by using our MEB quantizer. The CAP of our MEB (step 1 test) is consistently larger than 95\% when the number of tested IoT EDs grows from 0 to 200. However, more IoT EDs can lead to a higher probability of a failed step 1 test, which results in decreased CAPs when a large number of EDs appear. In this case, our step 2 test can further elevate the CAP as shown by the curves of MEB steps 1\&2 under both known/unknown $\sigma^2$.

According to Algorithm \ref{MEBalgorithm}, the MECD could offload the signing task only if our online two-step test authenticates the ED. Next, we evaluate the time efficiency improvement by the signing task offloading given by \eqref{partition}, \eqref{t1constrain}, \eqref{t2constrain}, and \eqref{constrain3}. Three different types of devices are considered to serve as either an IoT ED or an MECD. Specifically, we use the Texas Instrument Zigbee cc2430 node with a 32 MHz processor, which is commonly used for deploying IoT-based IEEE 802.15.4 wireless sensor networks. This device serves as the constrained ED. A Raspberry Pi 3 model-B with a 1.2 GHz CPU and a laptop with a 2.4 GHz CPU serve as MECD$_{P}$ and MECD$_{L}$, respectively. To compare the time efficiency, the same RSA and ECC signing algorithm is run by the three devices to obtain the fixed $\xi=\frac{B_sC_b}{f_{\text{ED/MEC}}}$ for each device. We assume that all NDN-IoT data packets' contents have been hashed to get the fixed-length digests, and thus the same $\xi$ is spent in signing on the digest for every NDN-IoT data packet. We show the time $N_p\xi$ needed by each type of device to complete the $N_p$ NDN-IoT packets' signings in Fig. \ref{RSAtime}, where $\phi=1, 0.3$ and $0.025$, respectively, indicate that 100\%, 30\%, and 2.5\% of the evaluation time of MECD$_{L/P}$ is available for the signing task offloading.
The optimal execution time of using \eqref{partition} is shown by MECD$_{L/P}$-ED (optim.). 
In Figs. \ref{RSAtime} (a) (b) (d) and (e), the execution time is reduced significantly since all $N_p$ packets are signed by either MECD$_{L}$ or MECD$_{P}$ instead of the constrained $\text{ED}$. In Fig. \ref{RSAtime} (c) and (f), the extremely busy MECD$_{L/P}$ ($\phi=0.025$) is not able to undertake all $N_p$ signings, and as a result the curves of MECD$_{L/P}$ become close to the curve of the IoT ED. In this case, our optimization offloading method can further reduce the execution time compared to letting either the MECD$_{L/P}$ or the IoT ED complete all tasks alone.

\subsection{Analysis of the Robustness Enhancement}\label{robustanalysis}

In this subsection, we analyze the robustness improvement gained through integrating the PHY-ID into the MEC-enabled NDN-IoT networks. It is considered that an attacker (ATTK) is trying to impersonate the legitimate $\text{ED}_A$ in order to injects his poisoned contents into NDN. 
This ATTK can be aware of the current pending interest for particular content or can anticipate the popular interests of consumers, and thus this ATTK is able to produce poisoned content under the popular name prefix and sign contents using the compromised key. Hence, once MECD/gateway is fooled, the poisoned content will be cached at NDN, and furthermore, copies of the poisoned content can be forwarded to any consumer with matched interests without triggering any alarm since the signature of ATTK is valid.

\textit{Attack 1:} It is assumed that an ATTK can act as the man-in-the-middle between ED$_A$ and MECD but does not know the authentication scheme. ATTK can passively eavesdrop on all NDN-IoT packets that are transmitted by the legitimate $\text{ED}_A$ without being detected. To mimic ED$_A$, this ATTK can send the intercepted packets to the MECD without any content modification (i.e., replay attacking). However, the device-specific IQI parameter, $a_{ATTK}$, that is associated with ATTK's RF hardware can inevitably affect the replayed packet. 
In our method, MECD can stealthily extract $a_{ATTK}$ from the replayed packets and generate $(\hat a_{ATTK}, q(\hat a_{ATTK}))_{\text{PHY-ID}_{ATTK}}$ rather than actively requesting that any pre-agreed secret be embedded into the NDN packet for authentication. Through checking the registered $(a_{A}, q(a_{A}))$, MECD can find $(\hat a_{ATTK}, q(\hat a_{ATTK}))\neq(a_{A}, q(a_{A}))$ and detect the presence of the ATTK since the ATTK's amplitude mismatch and phase-shift mismatch cannot be simultaneously the same as those of $\text{ED}_A$. Consequently, our method is secure against the replay attacks.

\textit{Attack 2:} It is considered that, occasionally, ATTK's  ${a}_{ATTK}$ is similar to ${a}_{A}$, which may challenge our authentication scheme if  $q(a_{A})=q(\hat a_{ATTK})$. Our offline/online phases are designed with a full consideration of this potential risk. At first, our offline MEB quantization technique ensures that ${a}_{ATTK}$ and ${a}_{A}$ can be quantized into different intervals with maximized probability. Then, as demonstrated in Fig. \ref{AuthImprovement}, our two-step authentication technique can accurately differentiate ${a}_{ATTK}$ and ${a}_{A}$ even if $q(a_{A})=q(\hat a_{ATTK})$. In addition, motivated by the MAMO technique \cite{IQIAFproof}, we find that exploiting more RF features, such as the carrier frequency offset (CFO), can further prevent this risk for two reasons: 1) taking more RF features into account can likely lead to sufficiently different ${a}_{ATTK}$ and ${a}_{A}$, and 2) our method that already uses two features (i.e., $\alpha$ and $\theta$) can be extended to incorporate more RF features. Taking the CFO as an example, we assume that \eqref{afunction} becomes $a=xg_3(\psi)+c$, where $ x=g_{1}(\alpha)g_2(\theta)$ and $\psi$ is the CFO parameter. After obtaining the PDF of $ x$ using \eqref{tau1}, $x g_3(\psi)+c$ can be processed as in the case of two features by implementing Algorithm \ref{case2algorithm}. Consequently, using our offline MEB quantization technique, online two-step authentication technique, and this potential MAMO technique together can make the NDN-IoT network robust to this kind of ATTK.

\textit{Attack 3:} We assume that the ATTK with strong computing power can compromise the signing key of ED$_A$, decipher our MEB quantization rule, and spoof and claim the legitimate $\text{PHY-ID}_A$ to the MECD. In this case, ATTK can be authenticated as legitimate by the traditional encryption-based authentication method since the signature can be created by ATTK using the compromised signing key of ED$_A$. In our method, the claimed $\text{PHY-ID}_A$ is not directly trusted by MECD but is used only as an index to find the registered $a_A$. In fact, MECD generates the $a_{ATTK}$ via physical-layer RF analysis. As long as $a_A\neq a_{ATTK}$, MECD can prevent ATTK with a high CAP and thereby refuse to send the poisoned contents to the NDN. If the claimed PHY-ID cannot point to a registered $a_A$, this ATTK's packets will be immediately rejected.
As a result, our MEC-enabled NDN-IoT network is secure against this computation-based PHY-ID spoofing attack.

In summary, Fig.~\ref{boundary}, Fig.~\ref{DiffRatefig}, and Fig. \ref{VB} validate the proposed MEB quantization technique and the two-step authentication technique. Fig. \ref{AuthImprovement} demonstrates the high authentication accuracy in the case of a large number of EDs. Fig. \ref{RSAtime} demonstrates the time efficiency enhancement using our signing task offloading. Finally, the analysis in Section \ref{robustanalysis} demonstrates the stronger robustness of our method gained by integrating the PHY-ID into the MEC-enabled NDN-IoT networks.

\section{Conclusion}\label{conclusionsection}

This paper investigated a design that integrates the PHY fingerprinting into the NDN signature scheme for the MEC-enabled NDN-IoT networks. With the aid of an MECD, the device-specific PHY-ID associated with the RF hardware of an IoT device can be generated using the proposed MEB quantization technique. This PHY-ID is then seamlessly integrated into the NDN signature with limited computational complexity imposed on the resource-constrained IoT devices. As a result, the MECD can examine IoT devices using our two-step authentication method and offload the signing task from the authenticated IoT device in an optimal manner. The evaluation results show that the MEC-enabled NDN-IoT networks can be improved in terms of the authentication accuracy, time efficiency, and robustness.

\appendices

\section{Proof for Theorem \ref{theorem1}}

Given $M$ quantization levels, the entropy $H$ can be maximized as
\begin{align}\label{maxentropy}
&\max_{\mathbf b} \left(H=-\sum_{m=1}^{M}\tilde f_A(a_{[m]})\log_2\left(\tilde f_A(a_{[m]})\right)\right)\end{align}
\begin{align}
&\text{s.t.} \ \ \sum_{m=1}^{M}\tilde f_A(a_{[m]})=\sum_{m=1}^{M}\left(\int_{b_m}^{b_{m+1}}f_A(x)\text dx\right)=1, \\
&\ \ \ \ \ \ \ \  a_{\text{min}}\leqslant b_0 \leqslant b_1 \leqslant \cdots\leqslant b_M\leqslant a_{\text{max}}
\end{align}where $\mathbf b=[b_0, b_1, \cdots, b_M]$. In fact, $\tilde f_A(a_{[m]})$ is dependent on $[a_{\text{min}}, a_{\text{max}}]$, $f_A(x)$, and $b_m$ in \eqref{quantization}, but only $b_m$ can be adjusted to maximize $H$. Setting the Lagrange multiplier variables as $\lambda$, \eqref{maxentropy} can be solved by
\begin{align}\label{Lagrange}   
&\frac{\partial \left(\sum\limits_{m=1}^{M}\tilde f_A(a_{[m]})\log_2(\tilde f_A(a_{[m]}))-\lambda(\sum\limits_{m=1}^{M} \tilde f_A(a_{[m]})-1)\right)}{\partial \tilde f_A(a_{[m]})}\nonumber\\
&=\sum\limits_{m=1}^{M}\log_2(\tilde f_A(a_{[m]}))+M\left(\frac{1}{\ln2}-\lambda\right)=0.
\end{align}Since every $\tilde f_A(a_{[m]})$ is only dependent on $\lambda$ and $\sum_{m=1}^M\tilde f_A(a_{[m]})=1$, we can conclude that $H$ can be maximized by setting $\tilde f_A(a_{[m]})=F_a(b_{m+1})-F_a(b_{m}) =\frac{1}{M}$ and $b_0=a_{\text{min}}$, which yields \eqref{MEBoundary}.

\section{Proof for Theorem \ref{theorem2}}
Using \eqref{afunction}, the key point of deriving $f_A(\cdot)$ becomes how to figure out the distribution of the product of $g_{1}$ and $g_{2}$. Note that the $\alpha$-related $g_{1}$ and the $\theta$-related $g_{2}$ are independent of each other. We define a transformation $\mathbb G_1=\{(g_1, g_2)|g_{\text{min1}}<g_1<g_{\text{max1}}, g_{\text{min2}}<g_2<g_{\text{max2}}\}\to \mathbb G_2=\{(x, g)|g_{\text{min1}}<x<g_{\text{max1}}, xg_{\text{min2}}<g<xg_{\text{max2}}\}$ as
\begin{align}\label{transformation}
&X=g_1,\ \ G=g_1g_2 .
\end{align}
Therefore, the inverse transformation function can be given by
\begin{align}\label{invTrans}
\mathbf f(x, g)=\left[x, \, \frac{g}{x}\right]^T,
\end{align}and the Jacobian of $\mathbf f(x, g)$ is
\begin{align}\label{jacobian}
J_{\mathbf f(x, g)}=
\begin{bmatrix}
 \frac{\partial \mathbf f(x, g)_{(1)}}{\partial x}&  \frac{\partial \mathbf f(x, g)_{(2)}}{\partial x}  \\ 
\frac{\partial \mathbf f(x, g)_{(1)}}{\partial g}  &  \frac{\partial \mathbf f(x, g)_{(2)}}{\partial g}  \\
\end{bmatrix}=
\begin{bmatrix}
 1 &  0  \\ 
 -\frac{g}{x^2}  &  \frac{1}{x}  \\
\end{bmatrix}.
\end{align}
The PDF of $g_1g_2$ can be computed based on the joint PDF of $X$ and $G$, $(x, g)\in\mathbb G_2$, as
\begin{align}\label{jointPDF}
\int_D f_{X, G}(x, g)\text dx&=\int_Df_{G_1}(\mathbf f(x, g)_{(1)}) f_{G_2}(\mathbf f(x, g)_{(2)})\nonumber \\
&\!\!\!\!\!\!\!\!\!\!\!\!\!\!\!\!\!\!\!\!\times\det (J_{\mathbf f(x, g)})\text dx
=\int_D f_{G_1}(x)f_{G_2}\left(\frac{g}{x}\right)\frac{1}{x}\text dx,
\end{align}where $D$ is the region $[g_{\text{min1}}, g_{\text{max1}}]$ with appropriate sub-intervals as shown in \eqref{tau1}. Finally, replacing $x$ with $x-c$ in \eqref{jointPDF} can yield the eq. \eqref{PDF1g}, \eqref{PDF3g} and \eqref{PDF2g} in Theorem \ref{theorem2}.

\section{Proof for Eq.~\eqref{GLRT} and \eqref{Fdist1}}
Without loss of generality, let $\mathbf y=[y_{[1]}, y_{[2]}, \cdots, y_{[N_s]}]^T$, and \eqref{offset} can be rewritten as $\mathbf y=\mathbf H_0\mathbf a_{\Delta}+\mathbf n_a$, where $\mathbf H_0$ is the $N_s\times p$ matrix of rank $p$, $\mathbf a_{\Delta}$ is a $p\times1$ offset vector, and  $\mathbf n_a$ is the $N_s\times1$ noise vector with PDF $\mathcal N(\bm0,\sigma^2\mathbf I_{N_s})$. We replace $\mathbf a_{\Delta}$ with $\mathbf A\mathbf a_{\Delta}$ in \eqref{binaryHypo}, where $\mathbf A$ is an $r\times N_s$ matrix of rank $r$. Then, the GLRT of $\mathbf y$ with unknown $\sigma^2$ is \cite[Eq. (9.14)]{Kay}
\begin{align}\label{GLRTRefer} 
\!\!\!\!\!\!L_{\text{GLRT}}&=\frac{N_s-p}{r}\frac{(\mathbf A\bm{\hat{a}}_{\Delta})^T(\mathbf A(\mathbf H_0^T\mathbf H_0)^{-1}\mathbf A^T)^{-1}(\mathbf A\bm{\hat{a}}_{\Delta})}{\mathbf y^T(\mathbf I_{N_s}-\mathbf H_0(\mathbf H_0^T\mathbf H_0)^{-1}\mathbf H_0^T)\mathbf y},
\end{align}where $\bm{\hat{a}}_{\Delta}=(\mathbf H_0^T\mathbf H_0)^{-1}\mathbf H_0^T\mathbf y$. Considering our system, $p=1, r=1, \mathbf A=1$ and $\mathbf H_0=[1,1,\cdots, 1]^T$ can be substituted into \eqref{GLRTRefer} to decide $\mathcal H_0/\mathcal H_0$ as
\begin{align}\label{GLRTDerive}
L_{\text{GLRT}}&\!\!=\!\!\frac{(N_s-1){\hat{a}}_{\Delta}^T(\mathbf H_0^T\mathbf H_0){\hat{a}}_{\Delta}}{\mathbf y^T\mathbf y-\mathbf y^T\mathbf H_0(\mathbf H_0^T\mathbf H_0)^{-1}\mathbf H_0^T\mathbf y}\nonumber \\
&\!\!=\!\!\frac{{(N_s-1)N_s\hat{a}}_{\Delta}^2}{\sum\limits_{k=1}^{N_s}y^2_{[k]}-\hat{a}_{\Delta}^T(\mathbf H_0^T\mathbf H_0)^{T}\hat{a}_{\Delta}}=\frac{N_s(N_s-1)\hat{a}_{\Delta}^2}{\sum\limits_{k=1}^{N_s}(y_{[k]}-\hat{a}_{\Delta})^2}
\mathop{\lessgtr}\limits_{\mathcal{H}_1}^{\mathcal{H}_0} b_{v},
\end{align}which confirms \eqref{GLRT}. Note that ${\hat{a}}_{\Delta}=\frac{1}{N_s}\sum_{k=1}^{N_s}y_{[k]}$ so that $\frac{1}{N_s}\sum_{k=1}^{N_s}y_{[k]}^2-\hat{a}_{\Delta}^2=\frac{1}{N_s}\sum_{k=1}^{N_s}y_{[k]}^2-2\hat{a}_{\Delta}(\frac{1}{N_s}\sum_{k=1}^{N_s}y_{[k]})+\hat{a}_{\Delta}^2=\frac{1}{N_s}\sum_{k=1}^{N_s}(y_{[k]}-\hat{a}_{\Delta})^2$ is used in deriving \eqref{GLRTDerive}.

Setting $x_1=\mathbf y^t \mathbf C \mathbf y=(\sqrt{N_s}\hat{a}_{\Delta}/\sigma)^2$, where $\mathbf C=\mathbf H_0(\mathbf H_0^T\mathbf H_0)^{-1}\mathbf H_0^T/\sigma^2$, and $x_2=\mathbf y^T\mathbf B\mathbf y$, where the rank of $\mathbf B=(\mathbf I_{N_s}-\mathbf H_0(\mathbf H_0^T\mathbf H_0)^{-1}\mathbf H_0^T)/\sigma^2$ is $N_s-1$, we can get $L_{\text{GLRT}}=\frac{x_1}{x_2/(N_s-1)}$. Since $\sqrt{x_1}\sim\mathcal{N}(\frac{\sqrt{N_s} a_{\Delta}}{\sigma}, 1)$, $x_1\sim \chi^2_1$ under $\mathcal H_0$ and $x_1\sim \chi'^2_1(\frac{N_sa^2_{\Delta}}{\sigma^2})$ under $\mathcal H_1$, where $ \chi^2_1$ and $\chi'^2_1(\frac{N_sa^2_{\Delta}}{\sigma^2})$ are the central/non-central chi-square distribution with 1 degree of freedom and $\frac{N_sa^2_{\Delta}}{\sigma^2}$ is the noncentrality parameter. $\mathbf B$ is idempotent as $\mathbf B^2=\mathbf B$. Since 
$\mathbf H_0^T\mathbf B=\mathbf B\mathbf H_0=\bm 0$, we obtain $x_2\sim\chi^2_{N_s-1}$ according to \cite[Eq.(2.29)]{Kay}. Since $\mathbf C (\mathbf \sigma^2\mathbf I_{N_s}) \mathbf B=\bm 0$, $x_1$ and $x_2$ are independent \cite{independence}. Given that the ratio of two independent $\chi^2$ random variables leads to an $F$-distributed random variable, the PDF of $L_{\text{GLRT}}$ is
\begin{align}\label{Fdist}
\!\!\!\!L_{\text{GLRT}}=\frac{x_1}{\frac{x_2}{N_s-1}}\!\!\sim\!\!\left\{
\begin{array}{clc}
&\!\!\!\!\!\!\!\!\!\!F(1,N_s-1) \ \ \ \ \ \ \ \ \ \ \, \text{under}\, \mathcal H_0\\
&\!\!\!\!\!\!\!\!\!\!F'(1,N_s-1,\frac{N_sa^2_{\Delta}}{\sigma^2}) \ \ \text{under}\, \mathcal H_1
\end{array}
\right.,
\end{align}which confirms eq. \eqref{Fdist1}.


\ifCLASSOPTIONcaptionsoff
  \newpage
\fi

\bibliographystyle{IEEEtran}
\bibliography{IEEEabrv,AFRef}

\end{document}